\documentclass[letterpaper, 10 pt, conference]{ieeeconf}  % Comment this line out if you need a4paper

\IEEEoverridecommandlockouts                              % This command is only needed if 
\usepackage{amsmath} % assumes amsmath package installed
\usepackage{amssymb}  % assumes amsmath package installed
\usepackage{mathrsfs} 
\usepackage{graphicx,cite}
\usepackage{enumerate}
\usepackage{comment}
\newtheorem{thm}{Theorem}

\newtheorem{cor}[thm]{Corollary}
\newtheorem{remark}{Remark}

\newtheorem{definition}{Definition}
\newtheorem{note}{Note}
\providecommand{\keywords}[1]{\textbf{\textit{Index terms---}} #1}
\title{\LARGE \bf
On Passivity, Feedback Passivity, And Feedback Passivity Over Erasure Network: A Piecewise Affine Approximation Approach
}

\author{Abhijit Mazumdar$^{1}$, Srinivasan Krishnaswamy$^{1}$ and Somanath Majhi$^{1}$% <-this % stops a space
\thanks{$^{1}$ Authors are with Department of Electronics \& Electrical Engineering,
        Indian Institute of Technology Guwahati, Guwahati-781039, India}
        \\{\tt\small \{abhijit.mazumdar, srinikris, smajhi\}@iitg.ac.in}%
}
\newcommand*{\QEDB}{\hfill\ensuremath{\square}}
\begin{document}

\maketitle
\thispagestyle{empty}
\pagestyle{empty}

%%%%%%%%%%%%%%%%%%%%%%%%%%%%%%%%%%%%%%%%%%%%%%%%%%%%%%%%%%%%%%%%%%%%%%%%%%%%%%%%
\begin{abstract}

In this paper, we deal with the problem of passivity and feedback passification of smooth discrete-time nonlinear systems by considering their piecewise affine approximations. Sufficient conditions are derived for passivity and feedback passivity. These results are then extended to systems that operate over Gilbert-Elliott type communication channels. As a special case, results for feedback passivity of piecewise affine systems over a lossy channel are also derived.

%In this letter, we first deal with the problem of passivity and feedback passification of a smooth discrete-time nonlinear system. We then go on to study the feedback passification over a Gilbert-Elliott type communication channel. Our approach is based on a piecewise affine approximation, where we partition the state-space into a number of polyhedral cells. In each polyhedral cell, the nonlinear system is approximated by a piecewise affine system. Sufficient conditions to ensure passivity, feedback passivity, and feedback passivity over a communication network are derived. Results for feedback passivity of a piecewise affine system over a Gilbert-Elliott type communication channel can also be derived easily as a special case.%Furthermore, it is shown that a piecewise affine state-feedback control laws are sufficient to make the closed-loop system passive with and without arbitrary packet losses. 

\end{abstract}
%%%%%%%%%%%%%%%%%%%%%%%%%%%%%%%%%%%%%%%%%%%%%%%%%%%%%%%%%%%%%%%%%%%%%%%%%%%%%%%%
%\vspace{-0.3cm}
\section{INTRODUCTION}
\label{intro} 
%\vspace{-0.2cm}
Passivity is an important tool to assess stability of nonlinear systems. In the recent past, passivity of smooth discrete-time nonlinear systems has been investigated in great detail  \cite{byrnes1991passivity,lin1995passivity,van2000l2,navarro2004feedback}. 
	\par Various problems relating to smooth nonlinear systems can be studied using a piecewise affine (PWA) approximation. %In fact, it is one of the main motivations for studying piecewise affine systems.
	The intuition for using such an approach stems from the fact that any smooth map can be locally approximated by an affine map with an arbitrary accuracy \cite{johansson1999piecewise}.
	\par Control of smooth nonlinear systems using  PWA approximations was introduced  in \cite{sontag1981nonlinear}. Therein, controllability and stabilization of PWA systems and nonlinear systems are studied. Dissipativity and passivity for a PWA system are studied in \cite{johansson1999piecewise} and \cite{iannelli2006complementarity,chang2007synthesis,bemporad2008passivity} respectively. %Further, \cite{johansson1999piecewise} investigates stability of a smooth autonomous nonlinear system using a PWA approximation.
	Passivity of a smooth nonlinear system by linearizing around the origin is studied in \cite{xia2014determining}. However, the results presented in \cite{xia2014determining} are valid only inside a small neighbourhood of the origin. 
	%\par To the best of our knowledge, passivity of nonlinear systems using a PWA approximation has not been investigated as yet. Also, controller design of nonlinear systems using the PWA approximation has not been addressed. 
	
	%The main contributions of this letter can be summarized as follows. We address three problems relating to  smooth nonlinear systems using a PWA approximation.	To start with, we derive conditions under which a smooth discrete-time nonlinear system becomes passive. Unlike \cite{xia2014determining}, our results are  valid for any neighbourhood (bigger or smaller) of the origin. Then, we derive conditions under which a piecewise linear state-feedback control law is sufficient to ensure feedback passivity. Finally, we consider the problem of feedback passivity of  smooth nonlinear systems over lossy communication network. One of the main features of the  approach presented in this letter is that it enables us to design the controllers
	%by solving certain linear matrix inequalities (LMIs) with some constraints. As LMIs are easy to solve, our approach provides a more efficient way of controller design.% in comparision to the approach used in earlier works such as \cite{navarro2004feedback}.   
	\par Systems whose subsystems, viz: controllers, actuators, and sensors are connected through a communication network are termed as networked control systems (NCSs).  Communication over a network results in data packet loss which can negitively impact the performance of a system\cite{schenato2007foundations}. Usually, an independent and identically distributed (i.i.d.) Bernoulli process or a Markov process is used to model packet losses. The Bernoulli process  model is used more often due to its mathematical tractability. However, packet loss in a realistic communication network could be temporally correlated.  Thus, a Markov process based model represents a more accurate model. Feedback passivity of a nonlinear system with packet losses has been studied in \cite{wang2014feedback}. 
	Their analysis is based on the frequency of packet losses that occur in the channel.
	\par The main contributions of this paper can be summarized as follows. We address three problems relating to  smooth nonlinear systems using a PWA approximation.	To start with, we derive conditions under which a smooth discrete-time nonlinear system becomes passive. Unlike \cite{xia2014determining}, our results are  valid for any neighbourhood (bigger or smaller) of the origin. Then, we derive conditions under which a piecewise linear state-feedback control law is sufficient to ensure feedback passivity. Finally, we consider the problem of feedback passivity of  smooth nonlinear systems over lossy communication network. Results corresponding to the feedback passivity of a PWA system with packet losses are also presented.
%	A special case of the final lemma presents a solution to the problem of feedback passivity of a PWA system with packet losses.
	One of the main features of the  approach presented in this paper is that it enables us to design the controllers
	by solving certain linear matrix inequalities (LMIs) which are subjected to additional constraints. As LMIs are easy to solve, our approach provides a more efficient way of controller design.
	%However, their analysis is not based on a appropriate packet loss model, and hence may not guarantee feedback passivity when subjected to realistic communication networks. 
	Unlike \cite{wang2014feedback}, we consider feedback passivity over a Gilbert-Elliott type communication channel where packet losses are modeled as a two-state Markov chain. Further, their approach is different than the one presented here.
	%Moreover, we investigate the problem using a piecewise affine approximation. 
	%\par To the best of the authors' knowledge, feedback passification of a PWA system over lossy communication channel has not been investigated so far. The solution to this problem is a special case of the final result presented in this letter.
	%Feedback passivity of a nonlinear system with packet losses has been studied in \cite{wang2014feedback}. However, a deterministic packet loss model is considered which does not represent the nature of packet losses that occur in a communication channel. We, on the other hand, consider feedback passivity over a Gilbert-Elliott type communication channel where packet losses are modeled as a two-state Markov chain. Moreover, unlike \cite{wang2014feedback}, we investigate the problem using a piecewise affine approximation. 
	\par Although there is literature available on the passivity of PWA systems, to the best of the authors' knowledge there is no literature that studies the passivity and feedback passivity of smooth nonlinear systems using a PWA approximation. Further, feedback passivity of a PWA system with packet loss has not been addressed as yet.
	\par The paper is structured as follows. Section II describes the problem along with some important notations.  Section III, contains the main results. In section IV, we demonstrate our results using a numerical example. Finally,  section V. presents the conclusion.\\
	%\textit{Notations:} \\
	%For a vector $v$, $||v||$ denotes the euclidean norm. $||G||$ denotes the induced $2$-norm of a matrix $G$. $\mathbb{E}$ represents mathematical expectation. $I$ denotes identity matrix of appropriate dimension. For a matrix $Y$, $Y\geq 0$, $Y\leq 0$, $Y>0$, and $Y<0$ denote that $Y$ is positive semidefinite, negative semidefinite, positive definite, and negative definite respectively. 
%\vspace{-0.6cm}	
\section{Problem Formulation} 
%\vspace{-0.15cm}
 %Let $(\Omega, \mathcal{F}, P)$ be the given probability space.
  Consider the smooth discrete-time nonlinear system:
  \small
  \begin{equation}
  \begin{split}
  x_{k+1}=f(x_k)+B_1u_k+D_1w_k\\
  z_k=h(x_k)+B_2u_k+D_2w_k,
  \end{split}
  \label{sys_eq_1}
  \end{equation}
  \normalsize
  wherein $x_k\in \mathbb{R}^n$ is the state vector, $u_k\in \mathbb{R}^m$ is the control input to the actuators, $w_k\in \mathbb{R}^s$ is the external input (or disturbance), $z_k\in \mathbb{R}^s$ is the output, $f:\mathbb{R}^n\rightarrow \mathbb{R}^n$, $h:\mathbb{R}^n\rightarrow \mathbb{R}^s$ are smooth maps, $B_1\in \mathbb{R}^{n\times m}$, $D_1\in \mathbb{R}^{n\times s}$, $B_2\in \mathbb{R}^{s\times m}$, $D_2\in \mathbb{R}^{s\times s}$ are constant matrices.  
  Many systems like power networks \cite{stegink2016unifying}, permanent magnet synchronous motors \cite{delpoux2014permanent}, and nonlinear RLC circuits \cite{jayawardhana2007passivity}, can be modeled by Equation (\ref{sys_eq_1}).
  \par As the maps $f(.)$ and $h(.)$ are smooth maps, it is possible to partition the state space into small regions and  approximate the nonlinear system (\ref{sys_eq_1}) in each region  by an affine system with arbitrary accuracy.  In particular, one can consider a polyhedral partition $\Big{\{}\mathscr{X}_i\Big{\}}_{i\in \mathcal{N}}$ of the state space, which is indexed by the set $\mathcal{N}$. $\mathcal{N}_0\subseteq \mathcal{N}$ denotes the index set for all cells that contains the origin while $\mathcal{N}_1\subseteq \mathcal{N}$ is the index set for the cells that do not contain the origin. Each polyhedral cell can be characterized by an equation of the form \cite{johansson1999piecewise}:
  \vspace{-0.2cm}
  \small
  \begin{equation}
  \begin{split}
  &{E}(i){x}_k+e(i) \geq 0   \hspace{0.2cm} x_k\in \mathscr{X}_i, \hspace{0.1cm} or \hspace{0.1cm} \bar{E}(i)\bar{x}_k\geq 0,
  \end{split}
  \label{partition}
  \end{equation}  
    \normalsize
  %This equation can be rewritten as,
  %\begin{equation}
  %\begin{split}
  %\textrm{Or}, \hspace{0.2cm}
  %\bar{E}(i)\bar{x}_k\geq 0,
  %\end{split}
  %\label{partition}
  %\end{equation}  
  where $\bar{E}(i):=[E(i) \hspace{0.3cm} e(i)]$, $\bar{x}(k):=[x(k) \hspace{0.3cm} 1]^T$. Thus, each element of the partition is characterised by the corresponding matrix $\bar{E}(i)$. Note that $e(i)=0$ if $i\in \mathcal{N}_0$.\\
  %We can also construct matrix $\bar{E}(i)=[E(i) \hspace{0.3cm} e(i)]$, with $e(i)=0$ if $i\in \mathcal{N}_0$, such that 
  %\begin{equation*}
  %\begin{split}
  %\bar{E}(i)\bar{x}_k \geq 0, \hspace{0.3cm} \textrm{if} \hspace{0.2cm} x_k\in \mathscr{X}_i
  %\end{split}
  %\end{equation*}  
  %Consider the following piecewise affine system:
  %\begin{equation}
  %\begin{split}
  %&{x}_{k+1}=A(i){x_k}+a(i)+B_1u_k+D_1w_k \hspace{0.3cm}\\
  %&{z}_k=C(i){x}_k+c(i)+B_2u_k+D_2w_k. \\
  %&  \textrm{for} \hspace{0.3cm} x_k\in \mathscr{X}_i
  %\end{split}
  %\label{sys_eq_2}
  %\end{equation}
  %where $a(i)=0$ and $c(i)=0$ if $i\in \mathcal{N}_0$.
\vspace{-0.5cm}
  \par   Let $\Big{\{}\mathscr{X}_i\Big{\}}_{i\in \mathcal{N}}$ be a polyhedral partition of the state space. % which is characterized by $\bar{E}(i)\bar{x}_k \geq 0, $. 
  Consider matrices $A(i)$, $C(i)$, and vectors $a(i)$, $c(i)$ for all $i\in \mathcal{N}$ such that,
  \small
  \begin{equation}
  \begin{split}
  &||f(x_k)-A(i)x_k-a(i)||=||m(x_k,i)||\leq \epsilon(i)||x_k||\\
  &||h(x_k)-C(i)x_k-c(i)||=||n(x_k,i)||\leq \delta(i)||x_k||\\ 
  \normalsize
  & \textrm{for}\hspace{0.1cm} x_k\in \mathscr{X}_i \hspace{0.1cm} \textrm{with} \hspace{0.1cm} \hspace{0.1cm} a(i)=0 \hspace{0.1cm} \textrm{and} \hspace{0.1cm} c(i)=0 \hspace{0.1cm} \textrm{if} \hspace{0.1cm} i\in \mathcal{N}_0 ,
  \end{split}
  \label{apprx}
  \end{equation}
  \normalsize
%\vspace{-0.05cm}
%We consider $||.||$ to be 2-norm for a vector and induced 2-norm for a matrix respectively. 
where, $m(x_k,i)$ and $n(x_k,i)$ are functions that contain the higher order terms of the functions $f(x_k)$ and $h(x_k)$, respectively.
 \par Now, with a polyhedral partition equipped with a PWA approximation as given by (\ref{apprx}), the nonlinear system (\ref{sys_eq_1}) can be expressed as follows:
  \small
  \begin{equation}
  \begin{split}
  &{x}_{k+1}= {A}(i){x}_k+a(i)+{B}_1u_k+{D}_1w_k+{m}(x_k,i)\\
  & z_k={C}(i){x}_k+c(i)+B_2u_k+D_2w_k+n(x_k,i),
  \end{split}
  \label{sys_eq_2}
  \end{equation}
  \normalsize
% Equation (\ref{sys_eq_2}) can be written in terms of the variable $\bar{x}_k$ as follows:
% If one assumes that $m(x_k,i)=0$ and $n(x_k,i)=0$, the nonlinear system (\ref{sys_eq_2}) becomes equivalent to the following PWA system:
%\small
 % \begin{equation}
 % \begin{split}
 % &{x}_{k+1}= {A}(i){x}_k+a(i)+{B}_1u_k+{D}_1w_k\\
 % & z_k={C}(i){x}_k+c(i)+B_2u_k+D_2w_k,
 % \end{split}
 % \label{sys_eq_pwa}
 % \end{equation}
  %\normalsize
 One can write (\ref{sys_eq_2}) in terms of the variable $\bar{x}_k$ as: 
  \small
  \begin{equation}
  \begin{split}
  &\bar{x}_{k+1}= \hat{A}(i)\bar{x}_k+\hat{B}_1u_k+\hat{D}_1w_k+\hat{m}(x_k,i)\\
  & z_k=\bar{C}(i)\bar{x}_k+B_2u_k+D_2w_k+n(x_k,i),
  \end{split}
  \label{sys_eq_nl}
  \end{equation}
  \normalsize
  where \small $\hat{A}(i):= \begin{aligned}
  \begin{bmatrix}
  A(i) \ $~~~~$ a(i)  \\
  0_{1 \times n} \ $~~~~$  1   
  \end{bmatrix} 
  \end{aligned}$, $\bar{C}(i):= \Big[C(i) \hspace{0.3cm} c(i)\Big]$,
  \\
  $\hat{B}_1(i):= \begin{aligned}
  \begin{bmatrix}
  B_1(i)   \\
  0_{1 \times m}    
  \end{bmatrix} 
  \end{aligned}$,
  $\hat{D}_1(i):= \begin{aligned}
  \begin{bmatrix}
  D_1(i)   \\
  0_{1 \times s}    
  \end{bmatrix} 
  \end{aligned}$, %$\hat{D}_2(i):= \begin{aligned}
  %\begin{bmatrix}
  %D_2(i)   \\
  %0    
  %\end{bmatrix} 
  %\end{aligned}$,
  \normalsize
  $\hat{m}(x_k,i):= \begin{aligned}
  \begin{bmatrix}
  m(x_k,i)   \\
  0    
  \end{bmatrix} 
  \end{aligned}$.
  \\ $0_{n\times m}$ denotes a $n\times m$ dimensional zero matrix.
  \normalsize \\
  %As all the states are accessible to the controller, the information it has at a time instant $k$ is given by the information set:
 % \small
 % \begin{equation*}
 % \mathcal{I}_k=\{x_0,x_1,...,x_k,v_0,v_1,...,v_{k-1}\}.
 % \end{equation*}   
  \normalsize
  Define passivity for nonlinear system (\ref{sys_eq_1}) with $u_k\equiv 0$ as: 
  \begin{definition}  \cite{byrnes1994losslessness}
  	The system (\ref{sys_eq_1}) with $u_k\equiv 0$ is said to be \textit{passive} if there exists a nonnegative function $V:\mathbb{R}^n\rightarrow \mathbb{R}^+$ with $V(0)=0$, called the \textit{storage function}, such that for all $x_k\in  \mathbb{R}^n$, $w\in  \mathbb{R}^s$, and for all $k\in \mathbb{Z}^+$:
  	\small
  	\begin{equation}
  	\begin{split}
  	V(x_{k+1})-V(x_k)\leq z_k^Tw_k.
  	\end{split}
  	\label{diss_ineq} 
  	\end{equation}
  	\normalsize
  	%If (\ref{diss_ineq}) holds with strict inequality, the system (\ref{sys_eq_1}) with $u_k\equiv 0$ is said to be strictly dissipative. \QEDB
  \end{definition}
  %\vspace{-0.2cm}
  \section{Main Results} 
  % Before going on to describe the main results, we state a few results related to vector norm and matrix induced norm which will be useful to this work. 
  %\par  Let $G^{n\times m}:\Big(\mathbb{R}^n, ||.||\Big) \rightarrow \Big(\mathbb{R}^m, ||.||\Big)$ be an linear operator on the normed vector space $\mathbb{R}^n$. The induced $2$-norm is defined as: 
  %\begin{equation*}
  %\begin{split}
  %||G||:= \underset{y\neq 0}{\textrm{sup}} \frac{||Gy||}{||y||}.
  %\end{split}
  %\end{equation*}
  %\vspace{-0.1cm}
  %In this section, we analyse passivity, controller design for feedback passivity and feedback passivity with control packet erasures. Further, in the sequel, we consider a polyhedral partition  $\Big{\{}\mathscr{X}_i \Big{\}}_{i\in \mathcal{N}}$ such that (\ref{apprx}) holds.
  %\vspace{-0.2cm}
 \subsection{Passivity and passification of the nonlinear system (\ref{sys_eq_1}):}
 %\vspace{-0.2cm}
	%We now derive conditions for the nonlinear system (\ref{sys_eq_1}) with $u_k\equiv 0$ to be passive.
	Consider a piecewise quadratic storage function of the form
	\small
	\begin{equation}
	\begin{split}
	V(x_k)&=  \frac{1}{2} \bar{x}_k^T \bar{P}(i) \bar{x}_k \hspace{0.1cm} \textrm{if} \hspace{0.2cm} x_k\in \mathscr{X}_i.
	\label{pp_1}
	\end{split}
	\end{equation}
	\normalsize
	If $i\in \mathcal{N}_0$ then $\bar{P}(i)$ is given by \small $\bar{P}(i)=diag\{P(i),0\}$. For each $i\in \mathcal{N}_0$, $P(i)$ is chosen such that $x_k^TP(i)x_k>0$ for all $x_k\neq 0$.  
	%If $i\in \mathcal{N}_0$ then  $\exists P(i) \in \mathbb{R}^{n \times n}$ such that \small $\bar{P}(i)=diag\{P(i),0\}$. \normalsize%, where $diag\{.\}$ denotes a block diagonal matrix. %$\bar{P}(i)=\left[\begin{matrix}P(i) & 0_{n \times 1}\\0_{1 \times n} &0 \end{matrix}\right]$. %where $0_n$ is a row vector with $n$ zeroes. 
	\normalsize
	\par Note that to find a nonnegative storage function $V(x_k)$ of the form given in (\ref{pp_1}), it is not necessary to look for matrices $\bar{P}(i)$, $i\in \mathcal{N}_1$, and ${P}(i)$, $i\in \mathcal{N}_0$, which are positive definite. Instead, matrices ${P}(i)$, $i\in \mathcal{N}_0$, and $\bar{P}(i)$, $i\in \mathcal{N}_1$ which result in ${x}_k^T {P}(i) {x}_k> 0$, when $x_k(\neq 0)\in \mathscr{X}_i$, $i\in \mathcal{N}_0$, and $\bar{x}_k^T \bar{P}(i) \bar{x}_k> 0$, when $x_k(\neq 0)\in \mathscr{X}_i$, $i\in \mathcal{N}_1$, respectively, will ensure nonnegativeness  of $V(x_k)$. This will enhance the flexibility in the selection of matrices ${P}(i)$.
	\par Following theorem presents conditions for passivity of system (\ref{sys_eq_1}) with $u_k\equiv 0$.
	%The following result presents a sufficient condition for the nonlinear system (\ref{sys_eq_1}) with $u_k\equiv 0$ to be passive with a piecewise quadratic storage function. 
	\vspace{-0.0cm}
	\begin{thm}
		%Consider the nonlinear system (\ref{sys_eq_1}) with a partition $\Big{\{}\mathscr{X}_i \Big{\}}_{i\in \mathcal{N}}$ such that (\ref{apprx}) holds. 
		The system (\ref{sys_eq_1}), with $u_k\equiv 0$, is passive if there exist symmetric matrices with positive entries $W(i)$, $R(i)$, and symmetric matrices $\bar{P}(i)$ such that the following inequalities are satisfied for all $i,j \in \mathcal{N}$: 
		\small
		\begin{equation}
		\begin{split}
		  \bar{P}(i)-\bar{E}^T(i)R(i)\bar{E}(i)>0, \hspace{0.2cm} \begin{aligned}
		\begin{bmatrix}
		\Lambda_{11}(i,j) $~~~~$  \Lambda_{12}(i,j)  \\
		\Lambda^T_{12}(i,j) \ $~~~~$  \Lambda_{22}(i,j)   
		\end{bmatrix} 
		\end{aligned}\leq 0,
		\end{split}
		\label{LMI_1}
		\end{equation}
		\normalsize
%\vspace{-0.2cm}
		\small
		\begin{equation*}
		\begin{split}
		\textrm{where}, \hspace{0.2cm}
		&\bar{P}(i)=\left[\begin{matrix}P(i) & 0_{n \times 1}\\0_{1\times n} &0 \end{matrix}\right]\hspace{0.2cm} \textrm{if} \hspace{0.2cm} i\in \mathcal{N}_0.\\
		%& 0_{n\times m} \hspace{0.1cm} \textrm{is a zero matrix of dimension} \hspace{0.1cm} n\times m\\
		&\Lambda_{11}= \hat{A}^T(i)\bar{P}(j)\hat{A}(i)-\bar{P}(i)+\bar{E}^T(i)W(i)\bar{E}(i)\\
		&\hspace{0.7cm}+ \Big(\rho_1(i,j)+\rho_2(i,j)+\rho_3(i,j)+\rho_4(i)\Big)I_{n+1} \\
		& \Lambda_{12}=\hat{A}^T(i)\bar{P}(j)\hat{D}_1(i)  - \bar{C}^T(i) \\
		& \hspace{-0.8cm} \Lambda_{22}= \Big(\hat{D}_1\Big)^T\bar{P}(j)\hat{D}_1-\Big[{D}^T_2+{D}_2\Big]+\Big(\rho_2(i,j)+\rho_4(i)\Big)I_s\\ 
		%	& \bar{P}(i)=\begin{aligned}
		%	\begin{bmatrix}
		%	P(i) $~~~~$  0  \\
		%	0 \ $~~~~~~$  0   
		%	\end{bmatrix} 
		%	\end{aligned}\\
		%& \theta(i,j)= \rho_1(i,j)+\rho_2(i,j)+\rho_3(i,j)+\rho_4(i)\\
		%& \Phi(i,j)=\rho_2(i,j)+\rho_4(i)\\
		& \hspace{-0.8cm} \rho_1(i,j)=2\epsilon(i) ||\bar{P}(j)||||\hat{A}(i)||, \hspace{0.2cm} \rho_2(i,j)=  {\epsilon(i) ||\bar{P}(j)||||\hat{D}_1||}\\
		& \rho_3(i,j)=  {\epsilon^2(i)||\bar{P}(j)||}, \hspace{0.2cm} \rho_4(i)= {\delta(i)}.\\
		\normalsize
		& \textrm{$I_n$ denotes a $n$-dimensional identity matrix}.
		% ||.|| \textrm{denotes both vector and matrix 2-norm}.
		%& \textrm{For a matrix $L$, $||L||$ is induced 2-norm.}
		\end{split}
		\end{equation*}
		\normalsize
	\end{thm}
	\textit{Proof:} With a polyhedral partition $\Big{\{}\mathscr{X}_i\Big{\}}_{i\in \mathcal{N}}$, one can find matrices ${E}(i)$ and $\bar{E}(i)$, for $i\in \mathcal{N}$, such that ${E}(i){x}_k\geq 0$ and $\bar{E}(i)\bar{x}_k\geq 0$. As a piecewise affine approximation is being employed, one can consider a piecewise quadratic storage function of the form $V(x_k)=\frac{1}{2} \bar{x}_k^T\bar{P}(i)\bar{x}_k$, for $x_k\in \mathscr{X}_i$, to study the passivity of for the system (\ref{sys_eq_1}).
	\\ The first inequality in (\ref{LMI_1}) ensures that $V(x_k)$ is positive definite. Further,
	%Hence, $V(x_k)$ is well suited to be a storage function. Consider the following:
	\small
	\begin{equation}
	\begin{split}
	&V(x_{k+1})-V(x_k)-z_k^Tw_k\\
	&=\frac{1}{2} \bar{x}_{k+1}^T\bar{P}(j)\bar{x}_{k+1} -\frac{1}{2} \bar{x}_k^T\bar{P}(i)\bar{x}_k-z_k^Tw_k\\
	&=\frac{1}{2} \Big[ \Big(\hat{A}(i)\bar{x}_k+\hat{D}_1w_k+\hat{m}(x_k,i)\Big)^T \bar{P}(j) \Big(\hat{A}(i)\bar{x}_k+\hat{D}_1w_k+\hat{m}(x_k,i)\Big) \\
	&-\bar{x}_k^T\bar{P}(i)\bar{x}_k-\Big(\bar{C}(i)\bar{x}_k+D_2w_k+n(x_k,i) \Big)^Tw_k\\
	&-w_k^T\Big(\bar{C}(i)\bar{x}_k+D_2w_k+n(x_k,i) \Big) \Big]\\
	&=\frac{1}{2} \Big[ \Big(\hat{A}(i)\bar{x}_k+\hat{D}_1w_k\Big)^T\bar{P}(j)(\hat{A}(i)\bar{x}_k+\hat{D}_1w_k)-\bar{x}_k^T\bar{P}(i)\bar{x}_k\\
	&- \Big(\bar{C}(i)\bar{x}_k+D_2w_k\Big)^Tw_k-w_k^T\Big(\bar{C}(i)\bar{x}_k+D_2w_k\Big)\\
	&+ 2 \bar{x}^T_k\hat{A}^T(i)\bar{P}(j)\hat{m}(x_k,i)+2w_k^T\hat{D}_1^T\bar{P}(j)\hat{m}(x_k,i)+ \hat{m}^T(i)\bar{P}(j)\hat{m}(x_k,i)\\
	&-2w_k^Tn(x_k,i)\Big].
	\end{split}
	\label{proof_1}
	\end{equation}
	\normalsize
	%It is easy to verify that:
	%\small
	%\begin{equation}
	%\begin{split}
	%& \Big(\bar{A}(i)\bar{x}_k+D_1w_k\Big)^TP(j)\Big(\bar{A}(i)\bar{x}_k+D_1w_k\Big)-x_k^TP(i)x_k\\
	%&= \Big(\hat{A}(i)\bar{x}_k+\hat{D}_1w_k\Big)^T\bar{P}(j)\Big(\hat{A}(i)\bar{x}_k+\hat{D}_1w_k\Big)-\bar{x}_k^T\bar{P}(i)\bar{x}_k.
	%\end{split}
	%\label{proof_1_1}
	%\end{equation}
	%\normalsize
%\vspace{-0.2cm}
	Using basic properties of $2$-norm and induced $2$-norm:
	\small
	\begin{equation}
	\begin{split}
	2\bar{x}_k\hat{A}^T(i)\bar{P}(j)\hat{m}(x_k,i) & \leq ||2\bar{x}_k\hat{A}^T(i)\bar{P}(j)\hat{m}(x_k,i)||\\
	& \leq 2||\bar{P}(j)||||\hat{A}(i)|| ||\bar{x}_k|| \epsilon(i) ||{x}_k|| \\
	& \leq \rho_1(i,j) ||\bar{x}_k||^2 \hspace{0.2cm} \textrm{as} \hspace{0.2cm} ||{x}_k||\leq ||\bar{x}_k||
	\end{split}
	\label{ineq_1}
	\end{equation}
	\begin{equation}
	\begin{split}
	2w_k^T\hat{D}_1^T\bar{P}(j)\hat{m}(x_k,i) & \leq ||2w_k^T\hat{D}_1^T\bar{P}(j)\hat{m}(i)||\\
	& \leq 2||\bar{P}(j)||||\hat{D}_1||||w_k||||\hat{m}(x_k,i)||\\
	& \leq  2 \epsilon(i)||\bar{P}(j)||||\hat{D}_1|| ||\bar{x}_k|| ||w_k||  
	\end{split}
	\label{ineq_2}
	\end{equation}
	\normalsize
	As $(||\bar{x}_k||-||w_k||)^2 \geq 0$: $2||\bar{x}_k||||w_k|| \leq  \Big(||\bar{x}_k||^2+||w_k||^2\Big).$ 
	%\begin{equation*}
	%2||\bar{x}_k||||w_k|| \leq  \Big(||\bar{x}_k||^2+||w_k||^2\Big).
	%\end{equation*}
	Thus, (\ref{ineq_2}) implies:
	\begin{equation}
	2w_k^T\hat{D}_1^T\bar{P}(j)\hat{m}(x_k,i) \leq \rho_2(i,j) \Big(||\bar{x}_k||^2+||w_k||^2\Big)
	\label{ineq_3}
	\end{equation}
	\begin{equation}
	\textrm{Similarly,} \hspace{0.2cm}
	\hat{m}^T(x_k,i)\bar{P}(j)\hat{m}(x_k,i)\leq  \rho_3 ||\bar{x}_k||^2
	\label{ineq_4}
	\end{equation}
	\begin{equation}
	-2w^T_kn(x_k,i) \leq \rho_4(i) \Big(||\bar{x}_k||^2+||w_k||^2\Big)
	\label{ineq_5}
	\end{equation}
	From (\ref{proof_1}), (\ref{ineq_1}), (\ref{ineq_3}), (\ref{ineq_4}), (\ref{ineq_5}), and using the fact that $W(i)$ has only positive elements:
	\begin{equation*}
	\begin{split}
	&V(x_{k+1})-V(x_k)-z_k^Tw_k\\
	& \leq \frac{1}{2}
	\begin{aligned}
	\begin{bmatrix}
	\bar{x}_k   \\
	w_k    
	\end{bmatrix} 
	\end{aligned}^T
	\begin{aligned}
	\begin{bmatrix}
	\Lambda_{11}(i,j) $~~~~$  \Lambda_{12}(i,j)  \\
	\Lambda^T_{12}(i,j) \ $~~~~$  \Lambda_{22}(i,j)   
	\end{bmatrix} 
	\end{aligned}
	\begin{aligned}
	\begin{bmatrix}
	\bar{x}_k   \\
	w_k    
	\end{bmatrix} 
	\end{aligned}
	\end{split}
	\end{equation*}
	If the second inequality in (\ref{LMI_1}) is satisfied for $i,j\in \mathcal{N}$, then:
	\begin{equation*}
	\begin{split}
	&V(x_{k+1})-V(x_k)\leq z_k^Tw_k
	\end{split}
	\end{equation*}
	Hence, the nonlinear system (\ref{sys_eq_1}), with $u_k\equiv 0$, is passive. \QEDB
	\begin{remark}
	Note that, for a specific $i\in \mathcal{N}$, the conditions given by (\ref{LMI_1}) need to be satisfied for all $j\in \mathcal{N}$. This is due to the fact that one may not have any knowledge about the external input $w_k$. Consequently, $x_{k+1}$ is unknown, and hence the cell where $x_{k+1}$ lies in is also unkown. \QEDB
	\end{remark}
	%\par %By the following lemma, we derive conditions under which the closed-loop system (\ref{sys_eq_nl}) with a piecewise linear state-feedback control law $u_k=K(i)x_k$ becomes passive.\\
	\par For a piecewise state feedback law $u_k = K(i)x_k$, where $i$ denotes the index of the state space partition, let $\mathscr{A}_{K}(i)$, $\bar{\mathscr{A}}_K(i)$ and  $\hat{\mathscr{A}}_K(i)$ denote the matrices 
	$A(i)+B_1(i)K(i)$, $\Big[\mathscr{A}_K(i) \hspace{0.3cm} a(i)\Big]$ and
	$\begin{aligned}
	\begin{bmatrix}
	\mathscr{A}_K(i) $~~$  a(i) \\
	0_{1\times n} $~~~~~$ 1
	\end{bmatrix} 
	\end{aligned}$ respectively. Let $\mathscr{C}_K(i)$ and $\bar{\mathscr{C}}_K(i)$ denote the matrices
	$ C(i)+B_2(i)K(i)$ and 
	$\begin{aligned}
	\begin{bmatrix}
	\mathscr{C}_k(i) $~~$  c(i) 
	\end{bmatrix},
	\end{aligned}$ respectively.
\par Now, we derive results for feedback passivity as follows.	
	\begin{thm}\label{lem_2}
		The nonlinear system (\ref{sys_eq_1}), with $u_k=K(i)x_k=W(i)U^{-1}(i)x_k$, is passive if there exist matrices ${T}(i)>0$, $U(i)$, $W(i)$, $R(i)>0$, $G(i)>0$ and  scalars $q,r,h> 0$ such that for all $i,j \in \mathcal{N}$
		\small
		\begin{subequations}
			\begin{equation}
			\begin{split}
			\hat{T}(j)=\begin{aligned}
			\begin{bmatrix}
			T(j) $~~~~$  0_{n\times 1}  \\
			0_{1\times n}  $~~~~~~~$  h 
			\end{bmatrix} 
			\end{aligned}>0,
			\end{split} 
			\label{lmi_0}
			\end{equation} 
			\begin{equation}
			\begin{split}
			\begin{aligned}
			\begin{bmatrix}
			\Omega_{11} (i) $~~~~$ 0_{n\times 1} $~~~~~~$  \Omega_{13} (i) $~~~~~~$ \Omega_{14} (i) $~~~~$ 0_{n\times 1} \\
			0_{1\times n} $~~~~~~$ \Omega_{22} $~~~~~~$ qc^T(i) $~~~~~~~$ qa^T(i) $~~~~$ q\\
			\Omega^T_{13} (i)  $~~~~$ qc(i) $~~~~~~~$  \Omega_{33}  $~~~~~~~~$   \Omega_{34} $~~~~~$ 0_{s\times 1}\\
			\Omega^T_{14} (i) $~~~$ qa(i) $~~~~~~~$ \Omega^T_{34}  $~~~~~~~~$  T (j) $~~~~$ 0_{n\times 1} \\
			0_{1\times n} $~~~~~~~$ q $~~~~~~~~~~$ 0_{1\times s} $~~~~~~$ 0_{1\times n} $~~~~~~$ {h}
			\end{bmatrix} 
			\end{aligned}\geq 0,
			\label{lmi_1_0}
			\end{split}
			\end{equation}
			\begin{equation}
			\begin{split}
			\Big[\gamma_1(i,j)+\gamma_2(i,j)+\gamma_3(i,j)+\gamma_4(i) \Big]I_{n+1} 
			\leq \mathscr{L}(i)
			%\begin{aligned}
			%\begin{bmatrix}
			%U^{-T}(i)R(i)U^{-1}(i) $~~~~$  0  \\
			%$~~~~~~~~$0  $~~~~~~~~~~~~~~~~~~~$  \frac{r}{q^2} 
			%\end{bmatrix} 
			%\end{aligned},
			\end{split}
			\label{ineq_p_0}
			\end{equation}
			\begin{equation}
			\Big[\gamma_2(i,j)+\gamma_4(i)\Big]I_{s} \leq G(i),
			\label{pppp}
			\end{equation}
			\normalsize
		\end{subequations}
		\normalsize
		\textrm{where}
		\small
		\begin{subequations}
			\begin{equation*}
			\Omega_{11}(i)= \Big[ U(i)+U^T(i)-T(i)\Big]-R(i)
			\end{equation*}
			\begin{equation*}
			\Omega_{13}(i)=U^T(i)C^T(i)+W^T(i)B_2^T
			\end{equation*}
			\begin{equation*}
			\Omega_{14}(i)=U^T(i)A^T(i)+W^T(i)B_1^T
			\end{equation*}
			\begin{equation*}
			\Omega_{22}=2q-(h+r), \hspace{0.1cm} \Omega_{33}= \Big(D_2^T+D_2\Big)-G(i), \hspace{0.2cm} \Omega_{34}=D_1^T.
			\end{equation*}
			%\begin{equation*}
			%\Omega_{34}=D_1^T.
			%\end{equation*}
			\begin{equation*}
			\gamma_1(i,j)= 2\epsilon(i) ||\bar{\mathscr{A}}_K(i)|| ||{T}^{-1}(j)||
			\end{equation*}
			\begin{equation*}
			\gamma_2(i,j)=  \epsilon(i) ||{T}^{-1}(j)||||D_1||
			\end{equation*}
			\begin{equation*}
			\gamma_3(i,j)=  \epsilon^2(i)||{T}^{-1}(j)||, \hspace{0.2cm} \gamma_4(i)= \delta(i). 
			\end{equation*}
			\begin{equation*}
	\mathscr{L}(i)=\begin{aligned}
	\begin{bmatrix}
	U^{-T}(i)R(i)U^{-1}(i) $~~~~$  0_{n\times 1}  \\
	$~~~~~$0_{1\times n}  $~~~~~~~~~~~~~~~$  \frac{r}{q^2}
	\end{bmatrix} 
	\end{aligned}.
	\end{equation*}
			%\begin{equation*}
			%\gamma_4(i)= \delta(i).
			%\end{equation*}
		\end{subequations}
		\normalsize
		%The controller gain at each $i\in \mathcal{N}$ is given by
	%	\small 
	%	\begin{equation*}
	%	K(i)= W(i)U^{-1}(i).
	%	\label{control_law_1}
	%	\end{equation*} 
	%	\normalsize
	\end{thm}
	\par \textit{Proof:}
	Suppose, LMIs (\ref{lmi_0}) and (\ref{lmi_1_0}) are satisfied. As all principal submatrices of a positive semidefinite symmetric matrix are also postive semidefinite, from (\ref{lmi_1_0}), one gets:
	\begin{equation*}
	\begin{split}
	& \hspace{1.5cm} \begin{aligned}
	\begin{bmatrix}
	\Omega_{11}(i) $~~~~$  0_{n\times 1}  \\
	0_{1\times n}  $~~~~~~~$  \Omega_{22} 
	\end{bmatrix} 
	\end{aligned} \geq 0\\
	& \implies \hat{U}(i) + \hat{U}^T(i) \geq \hat{T}(i)+ \begin{aligned}
	\begin{bmatrix}
	R(i) $~~~~$  0_{n\times 1}  \\
	0_{1\times n}  $~~~~$  r 
	\end{bmatrix} ,
	\end{aligned}
	\end{split}
	\end{equation*}
%\vspace{0.15cm}
	where, $\hat{U}(i):=diag{\{}U(i),q{\}}, \forall i\in \mathcal{N}$. \\
	\vspace{0.05cm}
	So, $\forall i\in \mathcal{N}$, one gets that $\hat{U}(i)$ is non-singular as $\hat{T}(i)>0$. The following inequality can be proved easily. 
	\small
	\begin{equation*}
	\hat{U}^T(i)\hat{T}^{-1}(i)\hat{U}(i) \geq  \Big[ \hat{U}(i) + \hat{U}^T(i) -\hat{T}(i) \Big],
	\end{equation*}
	\normalsize
	which implies:
	\small
	\begin{equation*}
	\begin{split}
	&{U}^T(i){T}^{-1}(i){U}(i)-R(i) \geq  \Big[ {U}(i) + {U}^T(i) -{T}(i) \Big]-R(i),\\
	& \textrm{and}  \hspace{0.2cm} \frac{q^2}{ h}-r \geq 2q-(h+r).
	\end{split}
	\end{equation*}
	\normalsize
	Therefore,
	\small
	\begin{equation}
	\begin{split}
	&\Omega' (i,j)\\
	&=\begin{aligned}
	\begin{bmatrix}
  	$~~$ \mathscr{O}(i) $~~~~~$ 0_{n\times 1} $~~~~~$  \Omega_{13} (i) $~~~~~~~$ \Omega_{14} (i) $~~~~~$ 0_{n\times 1} \\
	0_{1\times n} $~~~$ \frac{q^2}{h}- r $~~~$ qc^T(i) $~~~~~~$ qa^T(i) $~~~~$ q\\
	\Omega^T_{13} (i)  $~~~~$ qc(i) $~~~~~~~$  \Omega_{33}  $~~~~~~~~$   \Omega_{34} $~~~~~$ 0_{s\times 1}\\
	\Omega^T_{14} (i) $~~~$ qa(i) $~~~~~~~$ \Omega^T_{34}  $~~~~~~~~$  T (j) $~~~~$ 0_{n\times 1} \\
	0_{1\times n} $~~~~~~~$ q $~~~~~~~~$ 0_{1\times s} $~~~~~~~$ 0_{1\times n} $~~~~~~~$ {h} 
	\end{bmatrix} 
	\end{aligned}\geq 0,
	\label{lmi_1_0_1}
	\end{split}
	\end{equation}
	\normalsize
	where \small
	$\mathscr{O}(i)= \Big[ U^T(i)T^{-1}(i)U(i) \Big]-R(i)$. 
	\normalsize \\
	Define $\mathscr{P}(i)$ as $\mathscr{P}(i)=diag\Big{\{} U^{-1}(i),\frac{1}{q},I_s,I_n,1 \Big{\}}$. %where $I_n$ is $n$ dimensional identity matrix.
	Then from (\ref{lmi_1_0_1}):
	\small
	\begin{equation}
	\begin{split}
	& \mathscr{P}^T(i) \Omega' (i,j) \mathscr{P}(i) \geq 0 \\
	& \implies \begin{aligned}
	\begin{bmatrix}
	\hat{T}^{-1}(i)-\mathscr{L}(i) $~~~~~$  \bar{\mathscr{C}}_K^T(i) $~~~~~~~~~~$ \hat{\mathscr{A}}^T_K (i) \\
	\bar{\mathscr{C}}_K(i)   $~~~~~~~~~~~~~~$  \Omega_{33}  $~~~~~~~~~~~~$   \hat{D}_1^T\\
	 \hat{\mathscr{A}}_K (i)  $~~~~~~~~~~~~~~$ \hat{D}_1  $~~~~~~~~~~~~$  \hat{T}(j) 
	\end{bmatrix} 
	\end{aligned}\geq 0,
	\end{split}
	\label{ppp}
	\end{equation} 
	\normalsize
	%where
	%\small
	%\begin{equation*}
	%\mathscr{L}(i)=\begin{aligned}
	%\begin{bmatrix}
	%U^{-T}(i)R(i)U^{-1}(i) $~~~~$  0  \\
	%0  $~~~~~~~~~~~~~~~~~~~$  \frac{r}{q^2}
	%\end{bmatrix} 
	%\end{aligned}.
	%\end{equation*}
	%\normalsize
	As $\hat{T}(j)>0$, from (\ref{ppp}), we get the following inequality using Schur complement: 
	\small
	\begin{equation}
	\begin{split}
	\mathscr{S}(i,j)=\begin{aligned}
	\begin{bmatrix}
	\mathscr{S}_{11}(i,j) $~~~~~~~~$  \mathscr{S}_{12}(i,j)  \\
	\mathscr{S}_{12}^T(i,j)  $~~~~~~~~$  	\mathscr{S}_{22}(i,j) 
	\end{bmatrix}\leq 0
	\end{aligned},
	\end{split}
	\label{lmi_5_1}
	\end{equation}
	\normalsize
	%where 
	\small
	$\textrm{where}, \hspace{0.1cm}
		\mathscr{S}_{11}(i,j)= \hat{\mathscr{A}}_K^T(i)\hat{T}^{-1}(j) \hat{\mathscr{A}}_K(i)- \hat{T}^{-1}(i)+\mathscr{L}(i)$,\\
		$\mathscr{S}_{12}(i,j)=\hat{\mathscr{A}}_K^T(i)\hat{T}^{-1}(j)\hat{{D}}_1-\bar{\mathscr{C}}_K^T(i)$,\\
	$\mathscr{S}_{22}(i,j)=\hat{D}_1^T\hat{T}^{-1}(j) \hat{D}_1- (D_2^T+D_2)+G(i).$\\	
	%\begin{subequations}
	%	\begin{equation*}
	%	\textrm{where}, \hspace{0.2cm}
	%	\mathscr{S}_{11}(i,j)= \hat{\mathscr{A}}_K^T(i)\hat{T}^{-1}(j) \hat{\mathscr{A}}_K(i)- \hat{T}^{-1}(i)+\mathscr{L}(i)
	%	\end{equation*}
	%	\begin{equation*}
	%	\mathscr{S}_{12}(i,j)=\hat{\mathscr{A}}_K^T(i)\hat{T}^{-1}(j)\hat{{D}}_1-\bar{\mathscr{C}}_K^T(i)
	%	\end{equation*}
	%	\begin{equation*}
	%	\mathscr{S}_{22}(i,j)=\hat{D}_1^T\hat{T}^{-1}(j) \hat{D}_1- (D_2^T+D_2)+G(i).
	%	\end{equation*}
	%\end{subequations}
	\normalsize
	%From LMI (\ref{lmi_0}), one gets that  ${T}(i)>0$, hence, ${T}^{-1}(i)>0$ for all $i\in \mathcal{N}$.
	Now, consider a piecewise quadratic storage function of the form $V(x_k)=\frac{1}{2}x_k^T{T}^{-1}(i)x_k$ for $x_k\in \mathscr{X}_i$.  Substituting $u_k=K(i)x_k$ for $x_k\in \mathscr{X}_i$, and using (\ref{sys_eq_2}) we get:
	\small
	\begin{equation*}
	\begin{split}
	& V(x_{k+1})-V(x_k)-z_k^Tw_k\\
	&=\frac{1}{2} x_{k+1}^T{T}^{-1}(j)x_{k+1} -\frac{1}{2} x_k^T{T}^{-1}(i)x_k-z_k^Tw_k\\
	&=\frac{1}{2} \Big[ \Big(\bar{\mathscr{A}}_K(i)\bar{x}_k+D_1w_k+m(x_k,i)\Big)^T {T}^{-1}(j) \Big(\bar{\mathscr{A}}_K(i)\bar{x}_k+D_1w_k\\
	&+m(x_k,i)\Big) -x_k^T{T}^{-1}(i)x_k-\Big(\bar{\mathscr{C}}_K(i)\bar{x}_k+D_2w_k+n(x_k,i) \Big)^Tw_k\\
	&-w_k^T\Big(\bar{\mathscr{C}}_K(i)\bar{x}_k+D_2w_k+n(x_k,i) \Big) \Big]\\
	&=\frac{1}{2} \Big[ \Big(\bar{\mathscr{A}}_K(i)\bar{x}_k+D_1w_k\Big)^T{T}^{-1}(j)(\bar{\mathscr{A}}_K(i)\bar{x}_k+D_1w_k)\\
	&-x_k^T{T}^{-1}(i)x_k- \Big(\bar{\mathscr{C}}_K(i)\bar{x}_k+D_2w_k\Big)^Tw_k\\
	& -w_k^T\Big(\bar{\mathscr{C}}_K(i)\bar{x}_k+D_2w_k\Big) + 2 \bar{x}^T_k\bar{\mathscr{A}}_K^T(i){T}^{-1}(j)m(x_k,i)\\
	&  +2w_k^TD_1^T{T}^{-1}(j)m(x_k,i) +  m^T(x_k,i){T}^{-1}(j)m(x_k,i)\\
	&-2w_k^Tn(x_k,i)\Big].
	\end{split}
	\end{equation*} 
	\normalsize
	Similar to (\ref{ineq_1}), (\ref{ineq_3}), (\ref{ineq_4}), (\ref{ineq_5}), the following inequalities can be derived.\\
	\small
\hspace{1.3cm}	$2\bar{x}^T_k\bar{\mathscr{A}}_K^T(i){T}^{-1}(j)m(x_k,i) \leq  \gamma_1(i,j) ||\bar{x}_k||^2$,\\
 \hspace{1.3cm}	$2w_k^TD_1^T{T}^{-1}(j)m(x_k,i) \leq \gamma_2(i,j) \Big(||\bar{x}_k||^2+||w_k||^2\Big)$,\\
\hspace{1.3cm}	$m^T(x_k,i){T}^{-1}(j)m(x_k,i) \leq \gamma_3(i,j) ||\bar{x}_k||^2$,\\
 \hspace{1.3cm}	$-2w_k^Tn(x_k,i) \leq \gamma_4(i) \Big(||\bar{x}_k||^2+||w_k||^2\Big)$.
	\normalsize
\\	Also, it is easy to show that: \\
\small $\Big(\bar{\mathscr{A}}_K(i)\bar{x}_k+D_1w_k\Big)^T{T}^{-1}(j)(\bar{\mathscr{A}}_K(i)\bar{x}_k+D_1w_k)-x_k^T{T}^{-1}(i)x_k\\
	=\Big(\hat{\mathscr{A}}(i)\bar{x}_k+\hat{D}_1w_k\Big)^T\hat{T}^{-1}(j)(\hat{\mathscr{A}}(i)\bar{x}_k+\hat{D}_1w_k)-\bar{x}_k^T\hat{T}^{-1}(i)\bar{x}_k$.
	\normalsize
	Then, one gets the following:
	\small
	\begin{equation*}
	\begin{split}
	&V(x_{k+1})-V(x_k)- z_k^Tw_k \\
	&=\frac{1}{2} \begin{aligned}
	\begin{bmatrix}
	\bar{x}_k   \\
	w_k    
	\end{bmatrix} 
	\end{aligned}^T
	\begin{aligned}
	\begin{bmatrix}
	\mathscr{R}_{11}(i,j) $~~~~~~~~$  \mathscr{R}_{12}(i,j)  \\
	\mathscr{R}_{12}^T(i,j)  $~~~~~~~~$  	\mathscr{R}_{22}(i,j) 
	\end{bmatrix}
	\end{aligned}
	\begin{aligned}
	\begin{bmatrix}
	\bar{x}_k   \\
	w_k    
	\end{bmatrix} 
	\end{aligned},
	\end{split}
	\end{equation*}
	\normalsize
	\small
	where, \hspace{0.1cm}
	$\mathscr{R}_{11}(i,j)=\hat{\mathscr{A}}_K^T(i)\hat{T}^{-1}(j)\hat{\mathscr{A}}_K(i)-\hat{T}^{-1}(i)$\\
\hspace{2.5cm}	$+\Big(\gamma_1(i,j)+\gamma_2(i,j)+\gamma_3(i,j)+\gamma_4(i)\Big)I_{n+1}$,\\ 
\hspace{2.5cm} $\mathscr{R}_{12}(i,j)=\hat{\mathscr{A}}_K^T(i)\hat{T}^{-1}(j)\hat{D}_1-\bar{\mathscr{C}}_K^T(i)$,\\
$	\mathscr{R}_{22}(i,j)=\hat{D}_1^T\hat{T}^{-1}(j)\hat{D}_1-(D_2^T+D_2)+\big(\gamma_2(i,j)+\gamma_4(i)\big)I_s.$
	%\begin{equation*}
	%\begin{split}
 %\textrm{where}, \hspace{0.2cm}	\mathscr{R}_{11}(i,j)&=\hat{\mathscr{A}}_K^T(i)\hat{T}^{-1}(j)\hat{\mathscr{A}}_K(i)-\hat{T}^{-1}(i)\\
%	&+\Big(\gamma_1(i,j)+\gamma_2(i,j)+\gamma_3(i,j)+\gamma_4(i)\Big)I
%	\end{split}
%	\end{equation*}
%	\begin{equation*}
%	\mathscr{R}_{12}(i,j)=\hat{\mathscr{A}}_K^T(i)\hat{T}^{-1}(j)\hat{D}_1-\bar{\mathscr{C}}_K^T(i)
%	\end{equation*}
%	\begin{equation*}
%	\mathscr{R}_{22}(i,j)=\hat{D}_1^T\hat{T}^{-1}(j)\hat{D}_1-(D_2^T+D_2)+\big(\gamma_2(i,j)+\gamma_4(i)\big)I.
%	\end{equation*}
%	\normalsize
	Now, following (\ref{ineq_p_0}), (\ref{pppp}) and (\ref{lmi_5_1}), we get:
	\small
	\begin{equation*}
	V(x_{k+1})-V(x_k)-z_k^Tw_k\leq  \frac{1}{2} \begin{aligned}
	\begin{bmatrix}
	\bar{x}_k   \\
	w_k    
	\end{bmatrix} 
	\end{aligned}^T
	\mathscr{S}(i,j)
	\begin{aligned}
	\begin{bmatrix}
	\bar{x}_k   \\
	w_k    
	\end{bmatrix} 
	\end{aligned} 
	\leq 0
	\end{equation*}
	\normalsize
	Therefore, with $u_k=K(i)x_k$, the system (\ref{sys_eq_nl}) is passive.  \QEDB
	\begin{note}
	To get the desired $K(i)$, $T(i)$ in a specific cell, we first solve (\ref{lmi_1_0}). Then, it is checked whether the conditions (\ref{ineq_p_0}) and (\ref{pppp}) get satisfied. If these two conditions are satisfied, then the cell is a valid one. If they are not satisfied, we opt for a finer cell with smaller $\epsilon(i)$ and $\delta(i)$ such that the conditions get satisfied. \QEDB
	\end{note}
	%\vspace{-0.1cm}
	\subsection{\textbf{Feedback passivity over erasure network}}
	%\vspace{-0.15cm}
	In this section, we deal with the problem of feedback passification over a Gilbert-Elliott type communication channel. 
	\par Suppose $u'_k$ is the controller output and is sent to the actuators through a lossy network. Then, under the zero-input scheme \cite{schenato2009zero}, one can relate  $u_k$ (as defined in (\ref{sys_eq_1})) with $u'_k$ by the expression $u_k=v_ku'_k$,
	%\small
	%\begin{equation*}
	%u_k=v_ku'_k,
	%\end{equation*}
	%\normalsize
	where $v_k$ is binary random variable, and can either be $0$ or $1$. It represents the packet loss condition in the channel. At a time index $k$, $v_k=0$ ($v_k=1$) denotes a packet being lost (a successful packet delivery) from the controller side to the actuator side. 
%	\par Two protocols are used in the context of NCSs. One is TCP(Transmission Control Protocol)-like and the other one is UDP(User Datagram Protocol)-like. In TCP-like protocol, packet reception is acknowledged. However, there is no such acknowledgement in UDP-like protocol \cite{schenato2007foundations}. 
	\par In this work we consider a TCP-like protocol wherein packet reception is acknowledged. Under such a  protocol, with perfect state knowledge, one can define an information set given by: $\mathcal{I}_k=\{x_0,x_1,..,x_k,v_0,v_1,...,v_{k-1}\}$.
	\par  The Gilbert-Elliott type channel model is basically a two-state Markov chain $\{v_k\}$, where $v_k=0$ and $v_k=1$ represent the two states of the Markov chain. At a time stage $k\geq 1$, the packet arrival probabilities are given as: $Pr\big(v_k=1|v_{k-1}=0\big)=\alpha$ and $Pr\big(v_k=1|v_{k-1}=1\big)=1-\beta$. At $k=0$, packet arrival probabilities are given by: $Pr\big(v_0=1\big)=\alpha/(\alpha + \beta)$ and $Pr\big(v_0=0\big)=\beta/(\alpha + \beta)$ \cite{mo2013lqg}.

	%The state transition probability matrix of the Markov chain $\{v_k\}$ is given by: 
	%\small
	%\begin{equation*}
	%\begin{split}
	%&	\begin{aligned}
	%\begin{bmatrix}
	%Pr(v_k=0|v_{k-1}=0) \ $~~$ Pr(v_k=1|v_{k-1}=0)  \\
	%Pr(v_k=0|v_{k-1}=1) \ $~~$  Pr(v_k=1|v_{k-1}=1)   
	%\end{bmatrix} 
	%\end{aligned} \\
	%& =
	%\begin{aligned}
	%\begin{bmatrix}
	%1-\alpha \ $~~~~~~$ \alpha  \\
	%\beta \ $~~~~~~$  1-\beta   
	%\end{bmatrix} 
	%\end{aligned}.
	%\end{split}
	%\end{equation*}
	%\normalsize
	Consider a piecewise linear state-feedback control law $u'_k=K(i)x_k$, where $i$ is the index of the partition. Then, the nonlinear system (\ref{sys_eq_1}), with a polyhedral partition $\Big{\{}\mathscr{X}_i\Big{\}}_{i\in \mathcal{N}}$ satisfying (\ref{apprx}), takes the form:
	\small
	\begin{equation}
	\begin{split}
	&x_{k+1}=\Big(A(i)+v_kB_1K(i)\Big)x_k+a(i)+D_1w_k+m(x_k,i)\\
	&z_k=\Big(C(i)+v_kB_2K(i)\Big)x_k+c(i)+D_2w_k+n(x_k,i), \hspace{0.1cm} \textrm{if} \hspace{0.1cm} x_k\in \mathscr{X}_i.
%	& \textrm{if} \hspace{0.2cm} x_k\in \mathscr{X}_i.
	\end{split}
	\label{sys_cl_1}
	\end{equation}
	\normalsize
	Note that if one puts $m(x_k,i)\equiv 0$ and $n(x_k,i)\equiv 0$, for all $x_k$ and $i$, then the nonlinear system (\ref{sys_cl_1}) becomes equivalent to the following closed-loop PWA affine system with a piecewise linear state-feedback control law:
		\small
	\begin{equation}
	\begin{split}
	&x_{k+1}=\Big(A(i)+v_kB_1K(i)\Big)x_k+a(i)+D_1w_k\\
	&z_k=\Big(C(i)+v_kB_2K(i)\Big)x_k+c(i)+D_2w_k, \hspace{0.1cm} \textrm{if} \hspace{0.1cm} x_k\in \mathscr{X}_i.
%	& \textrm{if} \hspace{0.2cm} x_k\in \mathscr{X}_i.
	\end{split}
	\label{sys_cl_pwa}
	\end{equation}
	\normalsize
	\par Observe that due to the randomness of packet losses ($v_k$), the closed-loop systems (\ref{sys_cl_1}) and (\ref{sys_cl_pwa}) become stochastic in nature. Thus, to analyze the feedback passivity of the system (\ref{sys_cl_1}) and (\ref{sys_cl_1}), one needs a notion of stochastic passivity. Stochastic passivity, in the spirit of \cite{wang2013stochastic}, is defined as follows:  
	\begin{definition} 
		The system (\ref{sys_cl_1}) \big(similarly the system (\ref{sys_cl_pwa}) \big) is said to be \textit{passive} in the stochastic sense if there exists a nonnegative function $V:\mathbb{R}^n \times \mathcal{N} \rightarrow \mathbb{R}^+$, with $V(0,.)=0$, called the \textit{storage function}, such that for all $x_k\in  \mathbb{R}^n$, $w\in  \mathbb{R}^s$ and for all $k\in \mathbb{Z}^+$:
		\begin{equation*}
		\mathbb{E} \Big[V(x_{k+1},s_{k+1})\Big| \mathcal{I}_k\Big]-V(x_k,s_k) \leq \mathbb{E} \Big[z_k^Tw_k\Big| \mathcal{I}_k\Big],
		\end{equation*}
		where, $s_k \in \mathcal{N}$ denotes the cell in which $x_k$ lies in. \QEDB
	\end{definition}
    \par Following theorem presents results for feedback passivity with random packet losses.
	\begin{thm} 
		Consider the nonlinear system (\ref{sys_cl_1}) with given control packet arrival probabilities $\alpha$ and $1-\beta$. With a control law $u'_k=K(i)x_k=W(i)U^{-1}(i)x_k$, the nonlinear system  becomes stochastically passive if, for all $i,j\in \mathcal{N}$, there exist matrices $T(i)$, $U(i)$, $W(i)$, $R(i)>0$, $G(i)>0$, and positive scalars $h,r,q$ such that:
		\small
		\begin{subequations}
			\begin{equation}
			\begin{split}
			\hat{T}(j)=\begin{aligned}
			\begin{bmatrix}
			{T}(j) $~~~~$  0_{n\times 1}  \\
			0_{1\times n} \ $~~~~~~~$  h   
			\end{bmatrix} 
			\end{aligned}>0,
			\end{split} 
			\label{pp_0}
			\end{equation} 
			\begin{equation}
			\begin{split}
			\begin{aligned}
			\begin{bmatrix}
			\Omega_{11}(i) $~~$ 0_{n\times 1} $~~~~$  \Omega_{13} (i) $~~~$ \Omega_{14} (i) $~~~$ 0_{n\times 1} $~~~~$ \Omega_{16} (i) $~~$ 0_{n\times 1} \\
			 0_{1\times n} $~~~$ \Omega_{22}(i) $~~~$  qc^T(i) $~~$  qa^T(i) $~~~~$ q$~~~~~~~$ q a^T(i) $~~~$ q$~$\\
			\Omega^T_{13} (i)  $~~$  qc(i) $~~~~~$  \Omega_{33}  $~~~~~$   D^T_1 $~~~~~$ 0_{s\times 1} $~~~~~$   D^T_1 $~~~~$ 0_{s\times 1}\\
			\Omega^T_{14} (i) $~~$ q a(i) $~~~~~$ D_1  $~~~~~$  \Omega_{44} (j) $~~$ 0_{n\times 1} $~~~~$0_{n\times n} $~~~~$0_{n\times 1} \\
			 0_{1\times n} $~~~~~$ q $~~~~~$ 0_{1\times s} $~~~~~$ 0_{1\times n} $~~~~~$ \Omega_{55} $~~~~$0_{1\times n} $~~~~~~$0 \\
			\Omega^T_{16} (i) $~~$ q a(i) $~~~~~$ D_1 $~~~~~~~~$0_{n\times n} $~~~~$0_{n\times 1} $~~~$  \Omega_{66} (l) $~~$ 0_{n\times 1}  \\
			0_{1\times n} $~~~~~~$ q $~~~~~~~$ 0_{1\times s} $~~~~~$0_{1\times n} $~~~~~~$0 $~~~~~~$ 0_{1\times n} $~~~~$ \Omega_{77} 
			\end{bmatrix} \geq 0 
			\end{aligned} 
			\end{split}
			\label{lmi_1}
			\end{equation}
			\begin{equation}
			\begin{split}
			\Big[\rho_1(i,j)+\rho_2(i,j)+\rho_3(i,j)
			+\rho_4(i,l)+\rho_5(i,l)\\+\rho_6(i,l)+\rho_7(i) \Big]I_{n+1} \leq \mathscr{L}(i),
			\end{split}
			\label{ineq_p}
			\end{equation}
			\begin{equation}
			\Big[\rho_2(i,j)+\rho_5(i,l)+\rho_7(i)\Big]I_s\leq G(i),
			\label{ineq_pp}
			\end{equation}
		\end{subequations}
		\normalsize
		\textrm{where}
		\small
		\begin{subequations}
			\begin{equation*}
			\Omega_{11}(i)=   \Big[ U(i)+U^T(i)-T(i)\Big]-R(i)
			\end{equation*}
			\begin{equation*}
			\Omega_{22}= 2q-(h+r)    
			\end{equation*}
			\begin{equation*}
			\Omega_{13}(i)=U^T(i)C^T(i)+\bar{p}_kW^T(i)B_2^T
			\end{equation*}
			\begin{equation*}
			\Omega_{14}(i)=U^T(i)A^T(i)+W^T(i)B_1^T, \hspace{0.2cm} \Omega_{16}(i)=U^T(i)A^T(i)
			\end{equation*}
			\begin{equation*}
			\Omega_{33}=  \Big(D_2^T+D_2\Big)-G(i),  \Omega_{44}(j)= \frac{1}{\bar{p}_k} T(j), \hspace{0.2cm} \Omega_{55}= \frac{h}{\bar{p}_k}
			\end{equation*}
			\begin{equation*}
			\Omega_{66}(l)= \frac{1}{1-\bar{p}_k} T(l), \hspace{0.2cm} \Omega_{77}= \frac{h}{1-\bar{p}_k}
			\end{equation*}
			\begin{equation*}
			 \mathscr{L}(i) \hspace{0.1cm} \textrm{is as defined in Theorem 2},   
			\end{equation*}
			\begin{equation*}
			\rho_1(i,j)=2\bar{p}_k \epsilon(i) ||\bar{\mathscr{A}}_K(i)|| ||{T}^{-1}(j)||
			\end{equation*}
			\begin{equation*}
			\rho_2(i,j)= \bar{p}_k \epsilon(i) ||{T}^{-1}(j)||||D_1||
			\end{equation*}
			\begin{equation*}
			\rho_3(i,j)=\bar{p}_k \epsilon^2(i)||{T}^{-1}(j)|| 
			\end{equation*}
			\begin{equation*}
			\rho_4(i,l)=2(1-\bar{p}_k)\epsilon(i) ||\bar{A}(i)||||{T}^{-1}(l)||
			\end{equation*}
			\begin{equation*}
			\rho_5(i,l)=(1-\bar{p}_k)\epsilon(i)||{T}^{-1}(l)||||D_1||
			\end{equation*}
			\begin{equation*}
			\rho_6(i,j)=(1-\bar{p}_k) \epsilon^2(i)||{T}^{-1}(l)|| 
			\end{equation*}
			\begin{equation*}
			\rho_7(i)=\delta(i), \hspace{0.2cm} \bar{p}_k=
			\begin{cases}
			\alpha, \hspace{0.2cm} \textrm{if}\hspace{0.2cm} v_{k-1}=0\\
			1-\beta, \hspace{0.2cm} \textrm{if}\hspace{0.2cm} v_{k-1}=1
			\end{cases}
			\end{equation*}
		\end{subequations}
		%\item With known $\zeta,\eta$ and $\hat{T}(i)$ for all $i \in \mathcal{N}$ from the LMI (\ref{lmi_1}), the following is satisfied:
		%\small
		%\begin{equation}
		%\begin{split}
		%&\mathscr{G}(i,l)=\begin{aligned}
		%\begin{bmatrix}
		%	\mathscr{G}_{11}(i,l) $~~~~~~$   \mathscr{G}_{12}(i,l)  \\
		%\mathscr{G}^T_{12}(i,l)  $~~~~~~$  	\mathscr{G}_{22}(i,l) 
		%\end{bmatrix}\leq \begin{bmatrix}
		%	\hat{H}(i) $~~~~~~$   0  \\
		%0  $~~~~~~$  	M(i)
		%\end{bmatrix}
		%\end{aligned},
		%\end{split}
		%\label{lmi_6_1}
		%\end{equation}
		%\normalsize
		%where 
		%\begin{subequations}
		%\begin{equation*}
		%\mathscr{G}_{11}(i,l)=\Big(1-\bar{p}_k\Big)\hat{A}^T(i)\hat{T}^{-1}(l) \hat{A}(i)-(\xi-1) \hat{T}^{-1}(i)
		%\end{equation*}
		%\begin{equation*}
		%\mathscr{G}_{12}(i,l)=\Big(1-\bar{p}_k\Big)\hat{A}^T(i)\hat{T}^{-1}(l)\hat{D}_1
		%\end{equation*}
		%\begin{equation*}
		%\mathscr{G}_{22}(i,l)=\Big(1-\bar{p}_k\Big)\hat{D}_1^T\hat{T}^{-1}(l) \hat{D}_1- (q-1) \big(D_2^T+D_2 
		%\big).
		%\end{equation*}
		%\end{subequations}
		%\normalsize
	\end{thm}
	\par \textit{Proof:} Assume that LMIs (\ref{pp_0}) and (\ref{lmi_1}) are satisfied. Using the same line of argument as used in the proof for Theorem \ref{lem_2}, one gets that \small $\hat{U}(i):=diag{\{}U(i),q{\}}$ \normalsize is non singular. %Also,
	%\small
	%\begin{equation*}
	%{U}^T(i){T}^{-1}(i){U}(i) \geq  \Big[ {U}(i) + {U}^T(i) -{T}(i) \Big]
	%\end{equation*}
	%\begin{equation*}
	%\frac{q^2}{h}-r \geq 2q-(h+r).
	%\end{equation*}
	%\normalsize
	Consider the following:
	\small
	\begin{equation*}
	\begin{split}
	&\Omega'(i,j)=\\
	& \begin{aligned}
		\begin{bmatrix}
			\mathscr{O}_1(i) $~~~$ 0_{n\times 1} $~~~~$  \Omega_{13} (i) $~~~~$ \Omega_{14} (i) $~~~$ 0_{n\times 1} $~~~~$ \Omega_{16} (i) $~~$ 0_{n\times 1} \\
			 0_{1\times n} $~~~~$ \mathscr{O}_2 $~~~$  qc^T(i) $~~~~~$  qa^T(i) $~~~~$ q$~~~~~~~$ q a^T(i) $~~~$ q$~$\\
			\Omega^T_{13} (i)  $~~~$  qc(i) $~~~~~$  \Omega_{33}  $~~~~~~~$   D^T_1 $~~~~~$ 0_{s\times 1} $~~~~~$   D^T_1 $~~~~$ 0_{s\times 1}\\
			\Omega^T_{14} (i) $~~~$ q a(i) $~~~~~$ D_1  $~~~~~$  \Omega_{44} (j) $~~~~$ 0_{n\times 1} $~~~~$0_{n\times n} $~~~$0_{n\times 1} \\
			 0_{1\times n} $~~~~~~~$ q $~~~~~~~$ 0_{1\times s} $~~~~~$ 0_{1\times n} $~~~~~$ \Omega_{55} $~~~~$0_{1\times n} $~~~~~~$0 \\
			\Omega^T_{16} (i) $~~~$ q a(i) $~~~~~$ D_1 $~~~~~$0_{n\times n} $~~~$0_{n\times 1} $~~~~$  \Omega_{66} (l) $~~$ 0_{n\times 1}  \\
			0_{1\times n} $~~~~~~~~$ q $~~~~~~~$ 0_{1\times s} $~~~~~~~$0_{1\times n} $~~~~~$0 $~~~~$ 0_{1\times n} $~~~~~$ \Omega_{77} 
			\end{bmatrix}
	%\begin{bmatrix}
	%\mathscr{O}_1(i) $~~~~~$ 0 $~~~~~~$  \Omega_{13} (i) %$~~~~$ \Omega_{14} (i) $~~~~~$ 0 $~~~~$ \Omega_{16} (i) $~~~~~$ 0 \\
	%$~~~$ 0 $~~~~~~~$ \mathscr{O}_2 $~~~~~$  qc^T(i) $~~~~$  qa^T(i) $~~~~$ q$~~~~~$ q a^T(i) $~~~$ q\\
	%\Omega^T_{13} (i)  $~~~$  qc(i) $~~~~$  \Omega_{33}  $~~~~~~$   D^T_1 $~~~~~~~$ 0 $~~~~~~$   D^T_1 $~~~~~~$ 0\\
	%\Omega^T_{14} (i) $~~~$ q a(i) $~~~~~$ D_1  $~~~~$  \Omega_{44} (j) $~~~~$ 0 $~~~~~~~~$0 $~~~~~~~$0 \\
	%$~~~~$ 0 $~~~~~~~~$ q $~~~~~~~~~$ 0 $~~~~~~~~~$ 0 $~~~~~~$ \Omega_{55} $~~~~~$0 $~~~~~~~~$0 \\
	%\Omega^T_{16} (i) $~~~~$ q a(i) $~~~~~$ D_1 $~~~~~~~$0 $~~~~~~~$0 $~~~~~$  \Omega_{66} (l) $~~~~$ %0  \\
	%0 $~~~~~~~~~~~$ q $~~~~~~~~~$ 0 $~~~~~~~~~~$0 %$~~~~~~~$0 $~~~~~~$ 0 $~~~~~~$ \Omega_{77} 
	%\end{bmatrix} 
	\end{aligned}
	\label{lmi_3}
	\end{split}
	\end{equation*}
	\normalsize
	where 
	$\mathscr{O}_1(i)=  \Big[ U^T(i)T^{-1}(i)U(i) \Big]-R(i)$, $\mathscr{O}_2=\frac{q^2}{h}-r$.\\
	From (\ref{lmi_1}), using same reasoning as used in the proof for Theorem 2, it can be proved that $\Omega'(i,j)\geq 0.$\\
	Then, with $\mathscr{P}(i)=diag\Big{\{} U^{-1}(i),\frac{1}{q},I_s,I_n,1,I_n,1 \Big{\}}$,
	\small
	\begin{equation}
	\begin{split}
	&\mathscr{P}^T \Omega'(i,j) \mathscr{P} \hspace{0.2cm} \geq 0\\
	&Or 
	\begin{aligned}
	\begin{bmatrix}
	\hat{T}^{-1}(i)-\mathscr{L}(i) $~~~$  (\bar{\mathscr{C}}'_K(i))^T $~~~~~$ \hat{\mathscr{A}}_K^T (i) $~~~~~~~~~~$ \hat{{A}}^T (i) \\
	\bar{\mathscr{C}}_K'(i)   $~~~~~~~~~~$  \Omega_{33}  $~~~~~~~~~~~~~$   \hat{D}_1^T $~~~~~~~~~~$ \hat{D}_1^T\\
	\hat{\mathscr{A}}_K (i)  $~~~~~~~~~~~$ \hat{D}_1  $~~~~~~~~~~~~$  \Lambda_1(j) $~~~~~~~~~~$ \Lambda_3\\
$~~$	\hat{{A}} (i)  $~~~~~~~~~~~~$ \hat{D}_1 $~~~~~~~~~~~~~$ \Lambda_3 $~~~~~~~~~~~~$  \Lambda_2(l)
	\end{bmatrix} 
	\end{aligned}\geq 0,
	\label{lmi_cc}
	%\begin{aligned}
	%\begin{bmatrix}
	%\mathscr{G}(i) $~~~~~~$ 0_{n\times 1} $~~~~$  \mathscr{C}'_K^T(i) $~~$ \mathscr{A}_K^T (i) $~~$ 0_{n\times 1} $~~$ {A}^T (i) $~$ 0_{n\times 1} \\
	%0_{1\times n} $~~$ \frac{1}{h}-\frac{r}{q^2} $~~~$ c^T(i) $~~~~$ a^T(i) $~~~~$ 1 $~~~~$ a^T(i) $~~~~$ 1\\
	%\mathscr{C}'_K(i)  $~~~~~$  qc(i) $~~~~~$  \Omega_{33}  $~~~~~$   D^T_1 $~~~~$ 0_{s\times 1} $~~~$   D^T_1 $~~~$ 0_{s\times 1}\\
	%\mathscr{A} (i) $~~~~$ a(i) $~~~~~~$ D_1  $~~~~~~$  \Omega_{44} (j) $~$ 0_{n\times 1} $~$ 0_{n\times n} $~$ 0_{n\times 1} \\
	%0_{1\times n} $~~~~$ 1 $~~~~~~~~$ 0_{1\times s} $~~~~$ 0_{1\times n} $~~~~$ \Omega_{55} $~~~$ 0_{1\times n} $~~~$ 0$~~$\\
	%{A} (i) $~~~~$ a(i) $~~~~~~~$ D_1  $~~~~~$ 0_{n\times n}  $~~$ 0_{n\times 1}  $~~~$  \Omega_{66} (l) $~~$ 0_{n\times 1}\\
	%0_{1\times n} $~~~$ 1 $~~~~~~~~$ 0_{1\times s} $~~~~$ 0_{1\times n}  $~~~~$ 0 $~~~~~$ 0_{1\times n} $~~~$ \Omega_{77}
	%\end{bmatrix} 
	%\end{aligned}\geq 0
	%& \hspace{7cm} \geq 0
	%\label{lmi_4}
	\end{split}
	\end{equation}
	\normalsize
	%where, $\mathscr{G}(i)=  %T^{-1}(i)-U^{-T}(i)R(i)U^{-1}(i).$\\
	% LMI (\ref{lmi_4}) now transforms to:
	%\small
	%\begin{equation}
	%\begin{aligned}
	%\begin{bmatrix}
	%\hat{T}^{-1}(i)-\mathscr{L}(i) $~~~$  (\bar{\mathscr{C}}'_K(i))^T $~~~~~$ \hat{\mathscr{A}}_K^T (i) $~~~~~~~~~~$ \hat{{A}}^T (i) \\
	%\bar{\mathscr{C}}_K'(i)   $~~~~~~~~~~$  \Omega_{33}  $~~~~~~~~~~~~~$   \hat{D}_1^T $~~~~~~~~~~$ \hat{D}_1^T\\
	%\hat{\mathscr{A}}_K (i)  $~~~~~~~~~~~$ \hat{D}_1  $~~~~~~~~~~~~$  \Lambda_1(j) $~~~~~~~~~~$ \Lambda_3\\
%$~~$	\hat{{A}} (i)  $~~~~~~~~~~~~$ \hat{D}_1 $~~~~~~~~~~~~~$ \Lambda_3 $~~~~~~~~~~~~$  \Lambda_2(l)
%	\end{bmatrix} 
%	\end{aligned}\geq 0,
%	\label{lmi_cc}
%	\end{equation}
	\normalsize
	where,
	\small
	\begin{equation*}
	\begin{split}
	\Lambda_1(j) = \bar{p}^{-1}_k\hat{T}(j), \hspace{0.05cm} \Lambda_2(l) = \Big(1-\bar{p}_k\Big)^{-1}\hat{T}(l), \hspace{0.05cm} \Lambda_3=0_{(n+1)\times (n+1)},
	\end{split}
	\end{equation*}
	\begin{equation*}
	\mathscr{C}'_K(i)=C(i)+\bar{p}_k B_2(i)K(i), \hspace{0.2cm} \bar{\mathscr{C}}'_K(i)=\begin{aligned}
	\begin{bmatrix}
	\mathscr{C}'_K(i) $~~$  c(i) 
	\end{bmatrix}.
	\end{aligned}
	\label{cl_C}
	\end{equation*}
	\normalsize
	Using Schur complement, as $\hat{T}(j)>0$ for all $j\in \mathcal{N}$, (\ref{lmi_cc}) implies:
	\small
	\begin{equation}
	\begin{split}
	\begin{aligned}
	& \begin{bmatrix}
	\hat{\mathscr{A}}^T_K(i) $~~~~$  \hat{{A}}^T(i)  \\
	\hat{D}_{1}^T(i)  $~~~~$  	\hat{D}_{1}^T(i) 
	\end{bmatrix}
	\begin{bmatrix}
	\Lambda_1^{-1} $~~~~$  \Lambda_3  \\
	\Lambda_3  $~~~~$  	\Lambda_2^{-1} 
	\end{bmatrix}
	\begin{bmatrix}
	\hat{\mathscr{A}}_K(i) $~~~~$ \hat{D}_{1}(i)    \\
	\hat{{A}}(i)  $~~~~$  	\hat{D}_{1}(i) 
	\end{bmatrix} \\
	& - \begin{bmatrix}
	\hat{T}^{-1}(i)-\mathscr{L}(i) $~~~~$  (\bar{\mathscr{C}}'_K(i))^T  \\
	\bar{\mathscr{C}}'_K(i)  $~~~~~~~~~~~$  	\Omega_{33} 
	\end{bmatrix}
	\leq 0 \\
	& \implies \begin{bmatrix}
	\mathscr{S}_{11}(i,j) $~~~~~~~~$  \mathscr{S}_{12}(i,j)  \\
	\mathscr{S}_{12}^T(i,j)  $~~~~~~~~$  	\mathscr{S}_{22}(i,j) 
	\end{bmatrix}\leq 0
	\end{aligned},
	\end{split}
	\label{lem_p}
	\end{equation}
	\normalsize
	where, \\
	   \small $
		\mathscr{S}_{11}(i,j)=\bar{p}_k \hat{\mathscr{A}}^T_K(i)\hat{T}^{-1}(j) \hat{\mathscr{A}}_K(i)+(1-\bar{p}_k) \hat{{A}}^T(i)\hat{T}^{-1}(l) \hat{{A}}(i)\\
		 - \hat{T}^{-1}(i)+\mathscr{L}(i) $,
		 \\
		 $\mathscr{S}_{12}(i,j)=\bar{p}_k\hat{\mathscr{A}}_K^T(i)\hat{T}^{-1}(j)\hat{{D}}_1+(1-\bar{p}_k)\hat{{A}}^T(i)\hat{T}^{-1}(l)\hat{{D}}_1-(\bar{\mathscr{C}}'_K(i))^T$,
		 \\
		 $\mathscr{S}_{22}(i,j)=\bar{p}_k\hat{D}_1^T\hat{T}^{-1}(j) \hat{D}_1+(1-\bar{p}_k)\hat{D}_1^T\hat{T}^{-1}(l) \hat{D}_1- \Omega_{33}.$
	\normalsize
	%If the LMI (\ref{lmi_1}) is feasible, then there exist matrices $\hat{T}(i)>0$ for all $i\in \mathcal{N}$, so, $\hat{T}^{-1}(i)>0$. Hence, ${T}^{-1}(i)>0$. 
	Consider a piecewise quadratic storage function of the form $V(x_k,i)=\frac{1}{2}x_k^T {T}^{-1}(i)x_k$ if $x_k\in \mathscr{X}_i$. Assume that $x_{k+1}\in \mathscr{X}_j$ if $v_k=1$, and $x_{k+1}\in \mathscr{X}_l$ if $v_k=0$. Note that $s_{k+1}$ can either be $j$ or $l$ depending on $v_k$. Thus, with a control law $u_k=W(i)U^{-1}(i)x_k=K(i)x_k$ if $x_k\in \mathscr{X}_i$, $i\in \mathcal{N}$:
	\small
	\begin{equation*}
	\begin{split}
	&\mathbb{E}\Big[V(x_{k+1},s_{k+1})\Big| \mathcal{I}_k\Big]- V(x_k,i)-\mathbb{E}\Big[z_k^Tw_k \Big| \mathcal{I}_k\Big]\\
	&= \frac{1}{2}\Big{\{}\bar{p}_k \Big[\bar{\mathscr{A}}_K(i)\bar{x}_k+D_1w_k+m(x_k,i)\Big]^T{T}^{-1}(j)\\
	& \times \Big[\bar{\mathscr{A}}_K(i)\bar{x}_k+D_1w_k+m(x_k,i)\Big] \\
	& +  \Big(1-\bar{p}_k\Big) \Big[\bar{{A}}(i)\bar{x}_k+D_1w_k+m(x_k,i)\Big]^T{T}^{-1}(l)\\
	&\times \Big[\bar{A}(i)\bar{x}_k+D_1w_k+m(x_k,i)\Big] - x_k^T{T}^{-1}(i)x_k\\
	&-[\bar{\mathscr{C}}'_K(i)\bar{x}_k+D_2w_k+n(x_k,i)]^Tw_k-w_k^T[\bar{\mathscr{C}}'_K(i)\bar{x}_k+D_2w_k+n(x_k,i)]\Big{\}}
	\end{split}
	\end{equation*}
	
	%\vspace{0.01cm}
	\begin{equation*}
   \begin{split}
	&=\frac{1}{2} \Big{\{}\bar{p}_k \Big[\bar{\mathscr{A}}_K(i)\bar{x}_k+D_1w_k\Big]^T{T}^{-1}(j) \Big[\bar{\mathscr{A}}_K(i)\bar{x}_k+D_1w_k\Big] \\
	&+  \Big(1-\bar{p}_k\Big) \Big[\bar{{A}}(i)\bar{x}_k+D_1w_k\Big]^T{T}^{-1}(l) \Big[\bar{A}(i)\bar{x}_k+D_1w_k\Big]\\
	& - x_k^T{T}^{-1}(i)x_k-[\bar{\mathscr{C}}'_K(i)\bar{x}_k+D_2w_k]^Tw_k-w_k^T[\bar{\mathscr{C}}'_K(i)\bar{x}_k\\
	&+D_2w_k]+\bar{p}_k \Big[ 2 \bar{x}^T_k\bar{\mathscr{A}}_K^T(i){T}^{-1}(j)m(x_k,i)+2w_k^TD_1^T{T}^{-1}(j)m(x_k,i)\\
	&+  m^T(x_k,i){T}^{-1}(j)m(x_k,i)\Big]+(1-\bar{p}_k) \Big[ 2 \bar{x}^T_k\bar{{A}}^T(l){T}^{-1}(l)m(x_k,i)\\
	&+2w_k^TD_1^T{T}^{-1}(l)m(x_k,i)+  m^T(x_k,i){T}^{-1}(l)m(x_k,i)\Big]-2w_k^Tn(x_k,i)\Big{\}}
	\end{split}
	\end{equation*}
	\normalsize
	\begin{comment}
	Further,
	\small
	\begin{subequations}
		\begin{equation*}
		\begin{split}
		&\Big(\bar{\mathscr{A}}_K(i)\bar{x}_k+D_1w_k\Big)^T{T}^{-1}(j)(\bar{\mathscr{A}}_K(i)\bar{x}_k+D_1w_k)-x_k^T{T}^{-1}(i)x_k\\
		&=\Big(\hat{\mathscr{A}}_K(i)\bar{x}_k+\hat{D}_1w_k\Big)^T\hat{T}^{-1}(j)(\hat{\mathscr{A}}_K(i)\bar{x}_k+\hat{D}_1w_k)-\bar{x}_k^T\hat{T}^{-1}(i)\bar{x}_k,
		\end{split}
		\end{equation*}
		\normalsize
		and,
		\small
		\begin{equation*}
		2\bar{p}_k\bar{x}^T_k\bar{\mathscr{A}}_K^T(i){T}^{-1}(j)m(i) \leq  \rho_1(i,j) ||\bar{x}_k||^2
		\end{equation*}
		\begin{equation*}
		2\bar{p}_kw_k^TD_1^T{T}^{-1}(j)m(i) \leq \rho_2(i,j) \Big(||\bar{x}_k||^2+||w_k||^2\Big)
		\end{equation*}
		\begin{equation*}
		\bar{p}_km^T(i){T}^{-1}(j)m(i) \leq \rho_3(i,j) ||\bar{x}_k||^2
		\end{equation*}
		\begin{equation*}
		2(1-\bar{p}_k)\bar{x}^T_k\bar{{A}}^T(i){T}^{-1}(l)m(i) \leq  \rho_4(i,l) ||\bar{x}_k||^2
		\end{equation*}
		\begin{equation*}
		2(1-\bar{p}_k)w_k^TD_1^T{T}^{-1}(l)m(i) \leq \rho_5(i,l) \Big(||\bar{x}_k||^2+||w_k||^2\Big)
		\end{equation*}
		\begin{equation*}
		(1-\bar{p}_k)m^T(i){T}^{-1}(l)m(i) \leq \rho_6(i,l) ||\bar{x}_k||^2
		\end{equation*}
		\begin{equation*}
		-2w_k^Tn(i) \leq \rho_7(i) \Big(||\bar{x}_k||^2+||w_k||^2\Big)
		\end{equation*}
	\end{subequations}
	\normalsize
	\end{comment}
	Using the same line of argument as used in the proof for Theorem 2, we get:
	\small
	\begin{equation*}
	\begin{split}
	&\mathbb{E}\Big[V(x_{k+1},s_{k+1})\Big| \mathcal{I}_k\Big]- V(x_k,i)-\mathbb{E}\Big[z_k^T w_k \Big| \mathcal{I}_k \Big]\\
	&\leq \begin{aligned}
	\begin{bmatrix}
	\bar{x}^T_k $~~$ w^T_k    
	\end{bmatrix} 
	\begin{bmatrix}
	\mathscr{S}_{11}(i,j) $~~~~~~~~$  \mathscr{S}_{12}(i,j)  \\
	\mathscr{S}_{12}^T(i,j)  $~~~~~~~~$  	\mathscr{S}_{22}(i,j) 
	\end{bmatrix}
	\begin{bmatrix}
	\bar{x}_k \\
	w_k    
	\end{bmatrix} 
	\end{aligned}.
	\end{split}
	\end{equation*}
	\normalsize
	Now, from (\ref{lem_p}):
	\small
	\begin{equation*}
	\begin{split}
	& %\textrm{from (\ref{lem_p})}, \hspace{0.1cm}
	\mathbb{E}\Big[V(x_{k+1},s_{k+1})\Big| \mathcal{I}_k\Big]- V(x_k,i)-\mathbb{E}\Big[z_k^T w_k \Big| \mathcal{I}_k \Big] \leq 0.
	\end{split}
	\end{equation*}
	\normalsize
	Hence, closed-loop system (\ref{sys_cl_1}) is stochastically passive. \QEDB
	\par It is straight forward to see that, if one puts $\epsilon(i)=0$ and $\delta(i)=0$ in the above theorem, for all $i\in \mathcal{N}$, then the result corresponds to the result for feedback passivity of an PWA system over a Gilbert-Elliott type channel.
	\begin{cor}
Consider the PWA system given by (\ref{sys_cl_pwa}) with given control packet arrival probabilities $\alpha$ and $1-\beta$. The PWA system, with a piecewise linear state-feedback control law of the form $u'_k=K(i)x_k=W(i)U^{-1}(i)x_k$, becomes stochastically passive if, for all $i,j\in\mathcal{N}$, there exists matrices $T(i)>0$, $U(i)$, $W(i)$, and positive scalars $q,h$ such that:
\small
\begin{subequations}
 \begin{equation}
			\begin{split}
			\hat{T}(j)=\begin{aligned}
			\begin{bmatrix}
			{T}(j) $~~~~$  0_{n\times 1}  \\
			0_{1\times n} \ $~~~~~~~$  h   
			\end{bmatrix} 
			\end{aligned}>0,
			\end{split} 
			\label{pp_pwa}
			\end{equation} 
			\begin{equation}
			\begin{split}
			\begin{aligned}
			\begin{bmatrix}
			\Omega_{11}(i) $~~$ 0_{n\times 1} $~~~~$  \Omega_{13} (i) $~~~$ \Omega_{14} (i) $~~~$ 0_{n\times 1} $~~~~$ \Omega_{16} (i) $~~$ 0_{n\times 1} \\
			 0_{1\times n} $~~~$ \Omega_{22}(i) $~~~$  qc^T(i) $~~$  qa^T(i) $~~~~$ q$~~~~~~~$ q a^T(i) $~~~$ q$~$\\
			\Omega^T_{13} (i)  $~~$  qc(i) $~~~~~$  \Omega_{33}  $~~~~~$   D^T_1 $~~~~~$ 0_{s\times 1} $~~~~~$   D^T_1 $~~~~$ 0_{s\times 1}\\
			\Omega^T_{14} (i) $~~$ q a(i) $~~~~~$ D_1  $~~~~~$  \Omega_{44} (j) $~~$ 0_{n\times 1} $~~~~$0_{n\times n} $~~~~$0_{n\times 1} \\
			 0_{1\times n} $~~~~~$ q $~~~~~$ 0_{1\times s} $~~~~~$ 0_{1\times n} $~~~~~$ \Omega_{55} $~~~~$0_{1\times n} $~~~~~~$0 \\
			\Omega^T_{16} (i) $~~$ q a(i) $~~~~~$ D_1 $~~~~~~~~$0_{n\times n} $~~~~$0_{n\times 1} $~~~$  \Omega_{66} (l) $~~$ 0_{n\times 1}  \\
			0_{1\times n} $~~~~~~$ q $~~~~~~~$ 0_{1\times s} $~~~~~$0_{1\times n} $~~~~~~$0 $~~~~~~$ 0_{1\times n} $~~~~$ \Omega_{77} 
			\end{bmatrix} \geq 0 
			\end{aligned} 
			\end{split}
			\label{lmi_pwa}
			\end{equation}   
\end{subequations}
\normalsize
\textrm{where}
		\small
		\begin{subequations}
			\begin{equation*}
			\Omega_{11}(i)=    U(i)+U^T(i)-T(i)
			\end{equation*}
			\begin{equation*}
			\Omega_{22}= 2q-h   
			\end{equation*}
			\begin{equation*}
			\Omega_{13}(i)=U^T(i)C^T(i)+\bar{p}_kW^T(i)B_2^T
			\end{equation*}
			\begin{equation*}
			\Omega_{14}(i)=U^T(i)A^T(i)+W^T(i)B_1^T, \hspace{0.2cm} \Omega_{16}(i)=U^T(i)A^T(i)
			\end{equation*}
			\begin{equation*}
			\Omega_{33}=  D_2^T+D_2,  \Omega_{44}(j)= \frac{1}{\bar{p}_k} T(j), \hspace{0.2cm} \Omega_{55}= \frac{h}{\bar{p}_k}
			\end{equation*}
			\begin{equation*}
			\Omega_{66}(l)= \frac{1}{1-\bar{p}_k} T(l), \hspace{0.2cm} \Omega_{77}= \frac{h}{1-\bar{p}_k}.
			\end{equation*}
\end{subequations}
	\end{cor}
	\normalsize
	%\begin{remark}
	%	When $\epsilon(i)=0$ and $\delta(i)=0$, for all $i\in \mathcal{N}$, the results presented in Lemma 2 and Lemma 3 become equivalent to the results corresponding to feedback passification and feedback passification over lossy channel of PWA systems, respectively.
	%\end{remark}
	\vspace{0.2cm}
 \section{Numerical Example}
 %\vspace{-0.1cm}
	Consider the nonlinear system (\ref{sys_eq_1}) with the following system parameters:
	\small
	\begin{eqnarray*}
	\begin{split}
	& f(x_k)=[4sin(x^1_k)+x^2_k, \hspace{0.1cm} x^1_k+x^3_k, \hspace{0.1cm} x^1_k]', \hspace{0.2cm} B_1=[2, \hspace{0.1cm} 0 , \hspace{0.1cm} 1]', \\
	& D_1=[1, \hspace{0.1cm} 0.5, \hspace{0.1cm} 0]', \hspace{0.2cm} C=[1 \hspace{0.1cm} 0 \hspace{0.1cm} 0], \hspace{0.2cm} B_2=0.1, \hspace{0.2cm} D_2=2,
	\\
	%& \hspace{2cm} \textrm{where} \hspace{0.1cm} x_k=[x^1_k, \hspace{0.1cm} x^2_k, \hspace{0.1cm} x^3_k]'. 
	\end{split}
	\label{sys_1}
	\end{eqnarray*}
	\normalsize
	%\small
	%\begin{equation}
	%\begin{split}
	%& f_3(x^1_k)=\begin{cases} 1.3x^1_k-0.17, & \mbox{if} \hspace{0.2cm} 0.5 \leq x^1_k \\
	%sin(x^1_k), & \mbox{if}   \hspace{0.2cm} -0.5<x^1_k<0.5 
	%\\
	%1.3x^1_k+0.17, & \mbox{if}   \hspace{0.2cm} x^1_k < -0.5
	%\end{cases} 
	%\end{split}
	%\label{sys_2}
	%\end{equation}
	\normalsize 
	%External input is assumed to be  $w_k=2sin(0.24\pi k)exp(-k/25)$.
	\\\normalsize 
	where $x_k=[x^1_k, \hspace{0.1cm} x^2_k, \hspace{0.1cm} x^3_k]$. External input is assumed to be  $w_k=0.02sin(0.2\pi k)exp(-k/25)$.
\par
\begin{figure} [!htp]
		{\centering
			\includegraphics[scale=0.32]{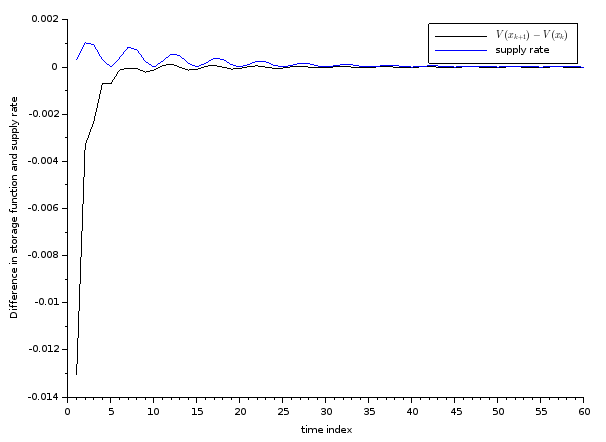} 
			\caption{ Difference in storage function and supply rate}
			\label{fig:Passivity}
		} 
	\end{figure}

	To demonstrate the results presented in Theorem 2, we calculate a piecewise linear state-feedback law that passifies the system in the region $-0.82\leq x^1_k \leq 0.82$  (note that nonlinearity exists only in $x_k^1$). The region $-0.82\leq x^1_k \leq 0.82$ is partitioned into $14$ cells, which are given by:
	$-0.82 \leq x^1_k<-0.78$, $-0.78 \leq x^1_k<-0.74$, $-0.74 \leq x^1_k<-0.7$, $-0.7 \leq x^1_k<-0.65$, $-0.65 \leq x^1_k<-0.6$, $-0.6 \leq x^1_k<-0.55$, $-0.55 \leq x^1_k<-0.5$, $-0.5 \leq x^1_k<-0.45$, $-0.45 \leq x^1_k<-0.4$, $-0.4 \leq x^1_k<-0.34$, $-0.34 \leq x^1_k<-0.28$, $-0.28 \leq x^1_k<-0.13$, $-0.13 \leq x^1_k<0$, $0 \leq x^1_k<0.13$, $0.13 \leq x^1_k<0.28$, $0.28 \leq x^1_k<0.34$, $0.34 \leq x^1_k<0.4$, $0.4 \leq x^1_k<0.45$, $0.45 \leq x^1_k<0.5$, $0.5 \leq x^1_k<0.55$, $0.55 \leq x^1_k<0.6$, $0.6 \leq x^1_k<0.65$, $0.65 \leq x^1_k<0.7$, $0.7 \leq x^1_k<0.74$, $0.74 \leq x^1_k<0.78$, $0.78 \leq x^1_k<0.82$. 
	Piecewise linear approximations are computed using the Taylor series expansion and the error term $\epsilon(i)$ is calculated in each cell.	
	  %Higher order terms are discarded and are replaced by an error term $m(i)$.\\
We use \textit{lmisolver} function available in SCILAB to solve the LMIs in each of the cells. %After solving LMI (\ref{lmi_1_0}), we check whether the conditions (\ref{ineq_p_0}) and (\ref{pppp}) are satisfied. If the conditions are not satisfied, we opt for a finer partition with smaller error $\epsilon(i)$ such that the conditions (\ref{ineq_p_0}) and (\ref{pppp}) get satisfied. 
	  Solving (\ref{lmi_1_0}) we then calculate the controller gain $K(i)$ such that the closed system becomes feedback passive.
	  The given cells are designed in such a way that the corresponding error term $\epsilon(i)$, in each $i\in \mathcal{N}$, is almost the maximum value that satisfies the conditions given by (\ref{ineq_p_0}) and (\ref{pppp}).   
	  From figure \ref{fig:Passivity}, one can see that difference in storage function at each stage is less than or equal to the supply rate. %External input is assumed to be  $w_k=0.02sin(0.24\pi k)exp(-k/25)$.
	 %For simulating feedback passivity with packet losses, we consider a nonlinear system where the function $f$ is as follows:  
%	\small
%	\begin{equation}
	%\begin{split}
	%&f(x_k)=[2x_k^1+x^2_k, \hspace{0.05cm} x_k^1+x_k^3, \hspace{0.05cm} f_3(x_k)]';\\
	%& f_3(x_k)=\begin{cases} 1.3x^1_k-0.1014, & \mbox{if} \hspace{0.2cm} 0.32 \leq x^1_k \\
	%sin(x^1_k), & \mbox{if}   \hspace{0.2cm} 0\leq x^1_k<0.32 
	%\\
	%x^1_k, & \mbox{if}   \hspace{0.2cm} x^1_k < 0
	%\end{cases}.
	%\end{split}
	%\label{sys_3}
	%\end{equation}
	%\normalsize 
 	\par
\begin{figure} [!htp]
		{\centering
			\includegraphics[scale=0.32]{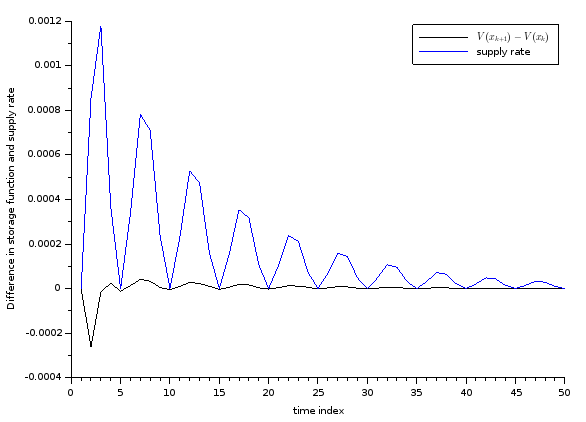} 
			\caption{Difference in storage function and supply rate}
			\label{fig:Passivity1}
		} 
	\end{figure}

 For feedback passivity with packet losses, we consider the region $-0.5\leq x_k^1 \leq 0.5$. The region $-0.5 \leq x_k^1 \leq 0$ is  partition  into cells: $-0.5\leq x_k^1 < -0.47$, $-0.47\leq x_k^1 < -0.44$, $-0.44\leq x_k^1 < -0.41$, $-0.41\leq x_k^1 < -0.38$, $-0.38\leq x_k^1 < -0.35$, $-0.35\leq x_k^1 < -0.32$, $-0.32\leq x_k^1 < -0.29$, $-0.29\leq x_k^1 < -0.26$, $-0.26\leq x_k^1 < -0.23$, $-0.23\leq x_k^1 < -0.2$, $-0.2\leq x_k^1 < -0.17$, $-0.17\leq x_k^1 < -0.14$, $-0.14\leq x_k^1 < -0.11$, $-0.11\leq x_k^1 < -0.07$, $-0.07\leq x_k^1 \leq 0$. The region $0\leq x_k^1 \leq 0.5$ is  partitioned in the similar fashion as the partition of the region $-0.5\leq x_k^1 \leq 0$, i.e., $0\leq x_k^1 < 0.07$, $0.07\leq x_k^1 < 0.11$ and so on. Then, solving LMI (\ref{lmi_1}) in Theorem 3, we calculate the controller gain $K(i)$ in each cells with a control packet arrival probability $\alpha=0.95$ and $1-\beta=0.96$. % With an external input $w_k=0.02sin(0.24\pi k)exp(-k/25)$,
 	Figure \ref{fig:Passivity1} demonstrates that the closed-loop system is passive with the quadratic storage function $V(x_k)=\frac{1}{2}x_k^T T^{-1}(i)x_k$. 
 	\par It is to be noted that, for feedback passivity with packet losses, we have to consider smaller cells as compared to feedback passivity. This is due to the fact that condition (\ref{ineq_p}) contains the term $\rho_4(i,l)$ which comes from open loop system dynamics.    
	
%State-response of the closed-loop system (\ref{sys_eq_1}) with parameters given by (\ref{sys_1}) and (\ref{sys_2}) and with control law $u_k=K(i)x_k$ is shown in Figure \ref{fig:stab_perfect}. Similarly, Figure \ref{fig:stab_ncs} shows the state response of the system (\ref{sys_cl_1})  with parameters given by (\ref{sys_1}) and (\ref{sys_3}). One can thus infer that the controllers (in Lemma 2 and Lemma 3) can stabilize the respective closed-loop systems with finite energy disturbance.
%\begin{figure} [!htp]
%	{\centering
%		\includegraphics[scale=0.4]{stability_perfect_channel.png} 
%		\caption{   State response of the closed-loop system with parameters given by (\ref{sys_1}) and (\ref{sys_2})}
%		\label{fig:stab_perfect}
%	} 
%\end{figure}
%\begin{figure} [!htp]
%	{\centering
%		\includegraphics[scale=0.4]{stability_ncs.png} 
%		\caption{   State response of the closed-loop system with parameters given by (\ref{sys_1}) and (\ref{sys_3})}
%		\label{fig:stab_ncs}
%	} 
%\end{figure}
h%\vspace{-0.2cm}
\section{CONCLUSIONS}
%\vspace{-0.2cm}
In this paper, using a piecewise affine approximation approach,  we have first derived sufficient conditions that ensures passivity of a smooth nonlinear system. Then, we have designed a piecewise linear state-feedback control law with which the closed-loop system becomes passive. Finally, with random control packet losses, a piecewise linear state-feedback control law is derived that makes the closed-loop system passive. Moreover, the problem of controller design for feedback passification of a PWA system over a lossy communication channel is also addressed as a special case of Theorem 3.

\addtolength{\textheight}{-12cm}   % This command serves to balance the column lengths
                                  % on the last page of the document manually. It shortens
                                  % the textheight of the last page by a suitable amount.
                                  % This command does not take effect until the next page
                                  % so it should come on the page before the last. Make
                                  % sure that you do not shorten the textheight too much.

%%%%%%%%%%%%%%%%%%%%%%%%%%%%%%%%%%%%%%%%%%%%%%%%%%%%%%%%%%%%%%%%%%%%%%%%%%%%%%%%

%%%%%%%%%%%%%%%%%%%%%%%%%%%%%%%%%%%%%%%%%%%%%%%%%%%%%%%%%%%%%%%%%%%%%%%%%%%%%%%%

%%%%%%%%%%%%%%%%%%%%%%%%%%%%%%%%%%%%%%%%%%%%%%%%%%%%%%%%%%%%%%%%%%%%%%%%%%%%%%%%

%%%%%%%%%%%%%%%%%%%%%%%%%%%%%%%%%%%%%%%%%%%%%%%%%%%%%%%%%%%%%%%%%%%%%%%%%%%%%%%%

%
%
%
%
%
%
%\end{thebibliography}
\vspace{-0.2cm}
\bibliographystyle{IEEEtran}
\bibliography{abhijit_csl_ref_1}

\end{document}